\documentclass[a4paper,fleqn,usenatbib,useAMS]{mnras}

\usepackage{graphicx}
\usepackage{lscape}
\usepackage{txfonts}
\usepackage{url}
\usepackage{lastpage}
\usepackage{gensymb}
\usepackage{supertabular}
\usepackage{longtable}
\usepackage{color}
\usepackage{soul}
\usepackage{hyperref}

\title[Spatial distribution of UDGs within large-scale structures]{Spatial distribution of ultra-diffuse galaxies within large-scale structures}

\author[Javier Rom\'an \& Ignacio Trujillo]{Javier Rom\'an$^{1,2}$ \thanks{E-mail:jroman@iac.es} and Ignacio Trujillo$^{1,2}$\\
$^{1}$Instituto de Astrof\'{\i}sica de Canarias, c/ V\'{\i}a L\'actea s/n, 
E-38205, La Laguna, Tenerife, Spain\\
$^{2}$Departamento de Astrof\'{\i}sica, Universidad de La Laguna, E-38206, La 
Laguna, Tenerife, Spain}

\begin{document}

\date{}

\pagerange{\pageref{firstpage}--\pageref{lastpage}} \pubyear{2016}

\maketitle

\label{firstpage}

\begin{abstract}

Taking advantage of the Sloan Digital Sky Survey Stripe82 data, we have explored the spatial distribution of ultra-diffuse galaxies (UDGs) within an area of 8$\times$8 Mpc$^2$ centred around the galaxy cluster Abell 168 ($z$ = 0.045). This intermediate massive cluster ($\sigma$ = 550 km s$^{-1}$) is surrounded by a complex large-scale structure. Our work confirms the presence of UDGs in the cluster and in the large-scale structure that surrounds it, and it is the first detection of UDGs outside clusters. Approximately 50 per cent of the UDGs analysed in the selected area inhabit the cluster region ($\sim$11 $\pm$ 5 per cent in the core and $\sim$39 $\pm$ 9 per cent in the outskirts), whereas the remaining UDGs are found outside the main cluster structure ($\sim$50 $\pm$ 11 per cent). The colours and the spatial distribution of the UDGs within this large-scale structure are more similar to the dwarf galaxies than to L$_\star$ galaxies, suggesting that most of UDGs could be bona fide dwarf galaxies.

\end{abstract}

\begin{keywords}

galaxies: dwarf -- galaxies: evolution -- galaxies: formation -- galaxies: photometry  -- galaxies: structure 

\end{keywords}


\section{Introduction}

Following pioneering work by \citet[][]{1988ApJ...330..634I}, \citet{1991ApJ...376..404B} and
\citet[][]{1997AJ....114..635D}, a number of recent works
\citep[e.g.][]{2015ApJ...798L..45V,2015ApJ...807L...2K,2015ApJ...809L..21M,2016A&A...590A..20V}, using deep imaging of nearby clusters, have focused their attention on a type of galaxy that has low mass (10$^7$-10$^8$ M$_\odot$), low surface brightness (24 $<$ $\mu_{g}(0)$ $<$ 26 mag arcsec$^{-2}$) and extended size (1.5 $<$ r$_e$ $<$ 4.5 kpc). \citet[][]{2015ApJ...798L..45V} have called these objects "ultra-diffuse galaxies" (UDGs). To date, UDGs have been found mostly in galaxy clusters. However, it is unclear whether this is just a bias produced by the strategy used to detect these objects. Because of their extreme low surface brightness, it is very time-consuming to obtain a spectroscopic redshift of these galaxies, and their redshifts have been estimated by their proximity to a high-density region. In fact, of the two UDGs confirmed spectroscopically \citep[][]{2015ApJ...804L..26V,2016AJ....151...96M}, one is in Coma and the other one is in a much lower density environment.  Moreover, there are also a small number of large, low surface brightness galaxies known in the field \citep[][]{1997AJ....114..635D,2001AJ....122.2318B}. None the less, despite the observational bias towards galaxy clusters,  the fact that the number of UDGs increases with cluster richness \citep[][]{2016A&A...590A..20V}, with almost a doubling of the population of known galaxies in Coma \citep[][]{2015ApJ...807L...2K} to barely adding a few candidates in Fornax \citep[][]{2015ApJ...813L..15M}, suggests that the environmental density could play a role in the origin of the UDGs. 

From the theoretical point of view,  \citet[][]{2015MNRAS.452..937Y} have shown that a scenario in which UDGs are satellites of a cluster, having  infallen early at $z$ $\sim$ 2 and quenching their further growth, is able to reproduce the structural properties of these objects. Interestingly, one of the predictions of this scenario is that  UDGs should not survive close to the centre of clusters, as tides exerted by the cluster mass within that region will disrupt the infalling UDGs. This seems to be in agreement with the findings of \citet[][]{2016A&A...590A..20V}. For all the above reasons, it is clear that understanding in which environments UDGs are originally formed, i.e. whether they have been formed in-situ in the clusters or whether they have been accreted through infalling substructures (group, filaments) is fundamental to disentangle the origin of these mysterious galaxies. Another key feature to better understand  the  nature of the UDGs is to compare their spatial distribution with two different families of galaxies: dwarf versus L$_\star$. Among other scenarios, \citet[][]{2015ApJ...798L..45V} discuss the intriguing possibility that
UDGs are failed L$_\star$ galaxies. Other works have explored the possibility that UDGs are regular dwarf galaxies  embedded in very massive dark matter halos \citep[e.g.][]{2016ApJ...819L..20B}, dwarf galaxies inhabiting high-spin halos \citep[][]{2016MNRAS.459L..51A}, failed Large Magellanic Cloud (LMC)-like galaxies \citep[][]{2016arXiv160408024B} or pure stellar halo galaxies \citep[][]{2016ApJ...822L..31P}.  We can shed extra light on the origin of these galaxies by addressing which spatial distribution UDGs resemble most.

In this paper, we explore the distribution and properties of the population of UDGs inside and around the Abell 168 galaxy cluster ($z$ = 0.045). Abell 168 is a cluster with a richness II-III (BM classification) located at RA(2000)=01h15m12.0s and Dec (2000)=+00d19m48s. This cluster has a velocity dispersion of $\sigma$ = 550 km s$^{-1}$ and a dynamical mass of M$_{dyn}$ = 5.2 $\times$ 10$^{14}$ M$_{\sun}$ \citep[][]{2004ApJ...600..141Y}. The large-scale structure surrounding this cluster is particularly relevant in order to probe where UDGs inhabit. In fact, there are many filaments and galaxy groups around Abell 168, as well as low-density regions. Our large-scale structure is fully embedded within the Sloan Digital Sky Survey (SDSS) Stripe 82. This means that we have deep photometry  in three SDSS bands ($g$ = 25.2 mag, $r$ = 24.7 mag and $i$ = 24.3 mag; 3$\sigma$ point sources), which implies having an extra filter compared to previous UDG works in other clusters. We will show that this extra filter is key to cleaning our galaxy sample from background  contaminants. This will allow us to select UDGs candidates at the redshift of the cluster, but significantly further away from its center. To characterize in great detail the large-scale structure around Abell 168, we use the deep spectroscopic coverage of this field produced by the SDSS survey in this area of the Stripe 82.  In this paper, we explore a field of view of 2.5\degree$ \times$ 2.5\degree, equivalent to 8 $\times$ 8 Mpc around this cluster.

The structure of the paper is as follows. In Section 2, we describe the  data. In Section 3, we explain how UDGs were identified. The structural properties of the UDG sample are detailed in Section 4 and their spatial distribution is explored in Section 5. In Section 6, we address a potential link between UDGs and regular dwarf galaxies. In Section 7, we present a discussion of the results, and finally, in Section 8, we summarize our main findings. Throughout this paper, we adopt the following cosmology $\Omega_m$=0.3, $\Omega_\Lambda$=0.7 and H$_0$=70 km s$^{-1}$ Mpc$^{-1}$. The spatial scale is 0.885 kpc \arcsec$^{-1}$ at the Abell 168 redshift ($z$ = 0.045). We use the AB magnitude system in this work.

\section{Data}

The images used in this paper were obtained from the Instituto de Astrof\'isica de Canarias (IAC) Stripe 82 Legacy Survey\footnote{\url{http://www.iac.es/proyecto/stripe82/}} \citep[][]{2016MNRAS.456.1359F}. The IAC Stripe 82 Legacy Survey consists of new deep co-adds of the Stripe 82 data from the SDSS, especially stacked to reach the faintest surface brightness limits of this data set. The average surface brightness limit is $r$ $\sim$ 28.5 mag arcsec$^{-2}$ (3$\sigma$ obtained in 10$\times$10 arcsec boxes). The Stripe 82 covers a 2.5 degree wide region along the celestial equator (-50\degree $<$ RA $<$ 60 \degree, -1.25 \degree $<$ Dec. $<$ 1.25 \degree) with a total of 275 deg$^{2}$. The pixel scale of the imaging is 0.396 arcsec. This region of the sky has been imaged repeatedly approximately 80 times in all the five SDSS filters (\textit{u, g, r, i} and \textit{z}). The IAC project provides the imaging dataset in 0.5$\times$0.5 degrees blocks. We created a final mosaic of 2.5$\times$2.5 degrees using SWarp\footnote{\url{http://www.astromatic.net/software/swarp}} \citep{2002ASPC..281..228B} centred approximately at the coordinates of the Abell 168 cluster (RA = 18.8\degree, Dec. = 0.33\degree).  The limiting surface brightness of the entire region was measured in each 0.5$\times$0.5 degrees block. We found that the surface brightness depth across the whole area is reasonably homogeneous, with a mean depth at the 3$\sigma$ level (in 10$\times$10 arcsec$^{2}$) of 29.2, 28.7 and 28.2 mag arcsec$^{-2}$ in the \textit{g, r} and \textit{i} filters respectively.

\section{Identification} \label{sec:sec2}

The goal of this paper is to find UDGs inside and around Abell 168. With this aim, we ran SExtractor \citep[][]{1996A&AS..117..393B} on the entire mosaic in the above 3 filters, plus another filter provided by the IAC project, \textit{r}-deep, which is the combination of \textit{g, r} and \textit{i}. As we are interested in  the detection of extended sources with very low surface brightness, we use a detection threshold in all the bands of 1$\sigma$ and a minimum area of 25 pixels (i.e. 4 arcsec$^2$ or 3 kpc$^2$ at the cluster redshift). With these settings, SExtractor found a total of 331 891 sources in the \textit{r}-deep filter. We further reduced our list of galaxies by requesting that the sources were identified in all the above bands. In addition, all the sources are selected to have simultaneously a stellarity (CLASS\_STAR) below 0.15 in each filter (to avoid selecting point-like objects).

The next selection criterion is based on the photometric colour (MAG\_AUTO) of the sources. We conservatively select the following ranges: 0 $<$ \textit{g-r} $<$ 1.2, 0 $<$ \textit{g-i} $<$ 1.7 and -0.2 $<$ \textit{r-i} $<$ 0.7. These colour intervals are broader than the expected colour values for the galaxy population (red and blue) at the cluster redshift (as we will show later). After all these selection criteria we are left with 75 666 galaxies.

The identification of UDGs requires a measurement of the size and surface brightness of the galaxies. For this reason, on all the-preselected galaxies we ran the IMFIT code \citep[][]{2015ApJ...799..226E}. We used a S\'ersic model \citep[][]{1968adga.book.....S} to extract the structural parameters of our pre-selected galaxies. We provided as input parameters for IMFIT the coordinates of the source, position angle and effective radius retrieved from the SExtractor run. IMFIT ran over the \textit{g, r} and \textit{i} filters. For each individual fitting, we masked all the pixels of nearby sources detected by SExtractor. The S\'ersic models were convolved with the point spread function (PSF) of the images. These PSFs are provided by the IAC Stripe 82 Legacy Survey for each individual block.

\begin{figure*}
  \centering
   \includegraphics[width=1.0\textwidth]{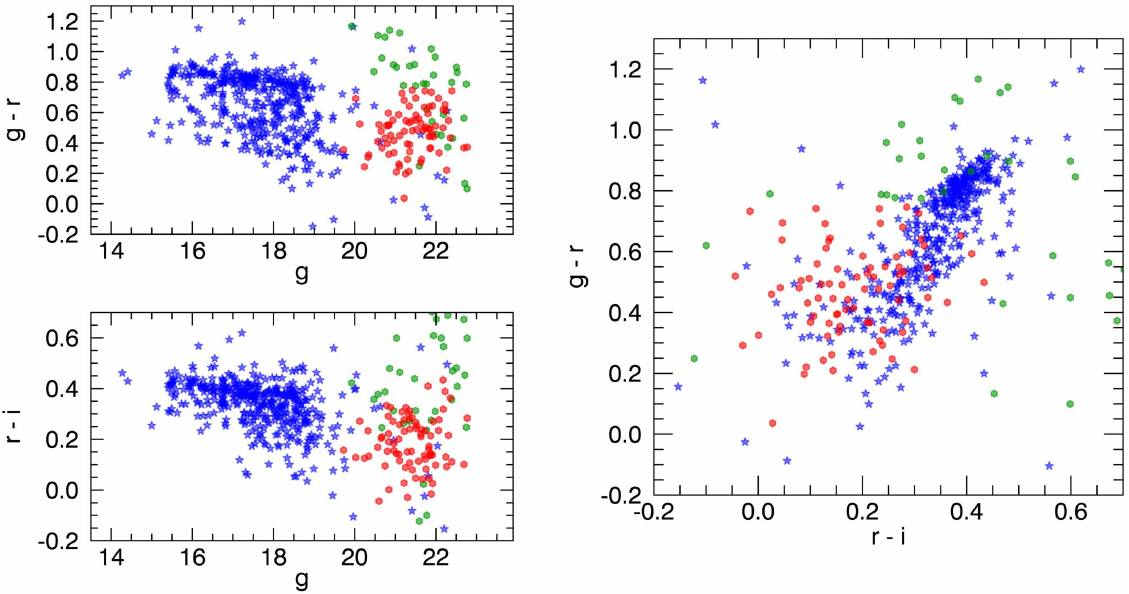}
   \caption{Colour-magnitude and colour-colour maps of the galaxies in the field of view of Abell 168.  The blue stars correspond to  galaxies in the area of Abell 168 with spectroscopic redshifts {0.037 $<$ z $<$ 0.052}. The red circles are the UDG candidates that have been selected as compatible with being at the cluster redshift using the
   colour distribution of the galaxies with spectroscopic redshifts. The green circles are those UDG candidates considered as potential contaminants because of their colours.}
 \label{fig:colores}
\end{figure*}

The structural parameters obtained from the S\'ersic fit are: position angle, ellipticity, S\'ersic index $n$, effective radius r$_e$ and total magnitude in each filter. Up to three slightly different input parameters in each individual fitting have been tried to ensure robustness of the output structural parameters of IMFIT. In most cases, the difference between the IMFIT results is negligible, but to be consistent we use the mean values of these three outputs. We compare the magnitude and effective radius values from IMFIT and SExtractor as a quality check of the fitting process, and we obtain a very good correlation between both magnitudes and a reasonably good correlation for the effective radii (see Fig. \ref{fig:appendixone}). The SExtractor effective radius value does not account for the PSF effect, so it is expected that these values will be slightly larger than those obtained by the IMFIT code, particularly for the most compact objects. Sources showing large magnitude differences ($>$1 mag) between IMFIT and SExtractor were flagged and visually inspected. These objects represent a tiny fraction of the total number of fitted sources ($\sim$0.5 per cent in the \textit{g} band). They are usually artifacts of the image, interacting galaxies, very bright galaxies with multiple pieces detected by SExtractor, bad fits of the IMFIT code, etc. We do not consider further these objects in the analysis.

\subsection{Selection of ultra-diffuse galaxies}\label{sec:ident}

Following previous works in the literature, the selection of UDGs in our field is based on the size and surface brightness of the galaxies. We use the structural parameters (magnitude, effective radius, $n$ and ellipticity) provided by IMFIT to perform our final cut. The magnitudes of the galaxies were corrected by the dust extinction of our Galaxy. Across our field of view, the dust extinction is relatively homogeneous and we select the following corrections: 0.098, 0.068 and 0.051 mag for the \textit{g, r} and \textit{i} filters, respectively \citep[][]{2011ApJ...737..103S}. Because of the low redshift of our galaxies, $z$ $\sim$ 0.045, the K-corrections are very small, K$_{g}$ $\sim$ 0.05, \citep[][]{2010MNRAS.405.1409C}\footnote{See, for example, http://kcor.sai.msu.ru/} and therefore we have not corrected our galaxies for this effect.

Assuming that UDGs are well described by an exponential light distribution, \citet[][]{2015ApJ...798L..45V} use the following criteria to select them: $\mu_{g}(0)>$ 24 mag/arcsec$^2$ and semimajor effective radius  r$_e>$ 1.5 kpc. By analyzing a large number of UDG candidates, \citet[][]{2015ApJ...807L...2K} show that their S\'ersic index $n$  distribution peaks towards slightly lower values than n = 1 (see their fig. 4). They also find that their axial ratio distribution has a broad shape with a maximum around 0.75. Based on this, we use the following criteria: 

\begin{enumerate} 

\item We use {$\mu_{g}(0)$ $>$ 24.0 mag arcsec$^{-2}$ after correction by dust extinction, where $\mu_{g}(0)$ is obtained using a S\'ersic model with $n$ free.}

\item We use circularized R$_e$ $>$ 1.25 kpc. Taking into account the typical axial ratio measured for these objects, this is equivalent to selecting galaxies with semimajor effective radius  r$_e$ $\gtrsim$ 1.5 kpc.

\end{enumerate}

After applying these criteria, we obtain 124 UDG candidates. All these candidates were visually inspected in order to eliminate artifacts misclassified as real UDGs. Examples of incorrectly detected UDGs are either groups of sources detected as a single object by SExtractor or mergers of galaxies. To visually discard some of these contaminants, we took advantage of colour stamps created by the combination of \textit{g, r} and \textit{i} filters. After this visual inspection we are left with 113 galaxies.

\begin{figure*}
  \centering
   \includegraphics[width=1\textwidth]{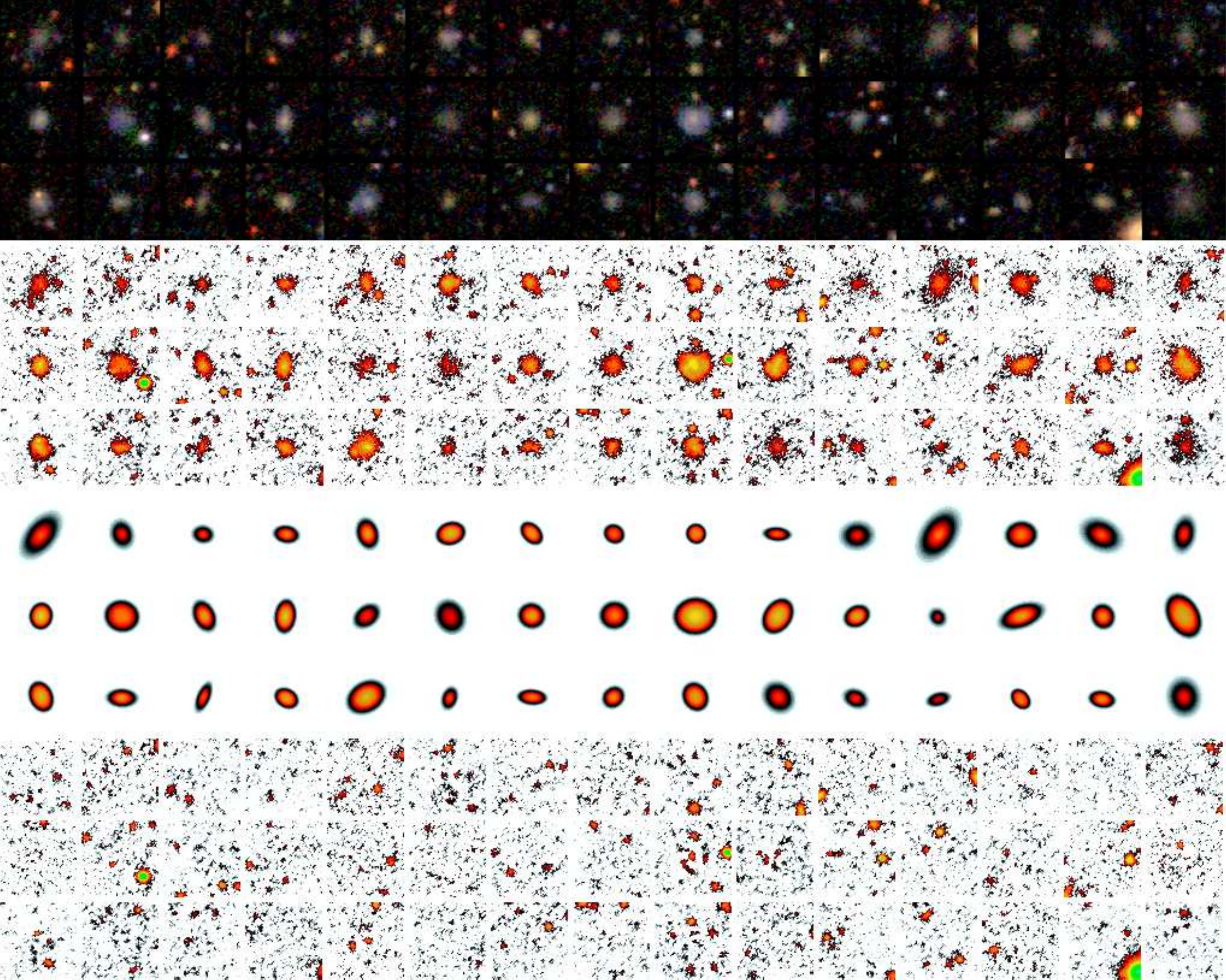}
   
   \caption{Mosaic showing a representative example of the UDGs explored in this paper. The colour stamps have been
created using the filters $g$, $r$ and $i$. The second row of images are the same galaxies in the $g$-band filter. The third row
corresponds to the IMFIT model used to parametrize the galaxies. The last set of images are the residuals after subtracting the IMFIT models from the images.}

   \label{fig:mosaico}
\end{figure*}

The next step is to select the UDGs that are at the distance of the Abell 168 structure and its surroundings. Lacking spectroscopic redshifts for our faint sources, the approach we have followed is to study the colour distribution of the galaxies with spectroscopic redshifts in the field compatible with being at the cluster redshift (i.e. $z$ = 0.045). In particular, we have selected all the galaxies with spectroscopic redshifts 0.037 $<$ $z$ $<$ 0.052 (i.e. those compatible with being within 3$\sigma$ of the velocity distribution of the Abell 168 cluster and the observed redshift distribution around this peak in the redshift histogram). The colour-magnitude and colour-colour maps for these spectroscopic sources are presented in Fig. \ref{fig:colores}. The limiting apparent magnitude in this field for having spectroscopic redshift is \textit{g} $\sim$ 19 mag. Fig. \ref{fig:colores} shows that  galaxies with spectroscopic redshifts have a broad distribution in colour, ranging from red galaxies defining a clear red sequence to objects following a blue cloud. The individual error in magnitude for each of the sources is key to characterize the broadening of the colour distribution\footnote{The magnitude uncertainty for each of our UDG candidates is $\sigma_g$ = 0.07 mag, $\sigma_r$ = 0.07 mag and $\sigma_i$ = 0.11 mag. These values are obtained comparing the magnitudes obtained in IMFIT versus SExtractor for galaxies in the range 20.5 $<$ g $<$ 22.5.}, particularly in the colour-colour maps. We have used the regular SDSS photometric data for our spectroscopic sources (this data is $\sim$2 mag shallower than Stripe 82). By doing this, we obtain (for the faintest galaxy with spectroscopic redshift) an uncertainty in the colour-colour map that is close to the one we obtain for the $\sim$2 mag fainter UDGs using Stripe 82 data. 

\begin{figure*}
  \centering
   \includegraphics[width=0.75\textwidth]{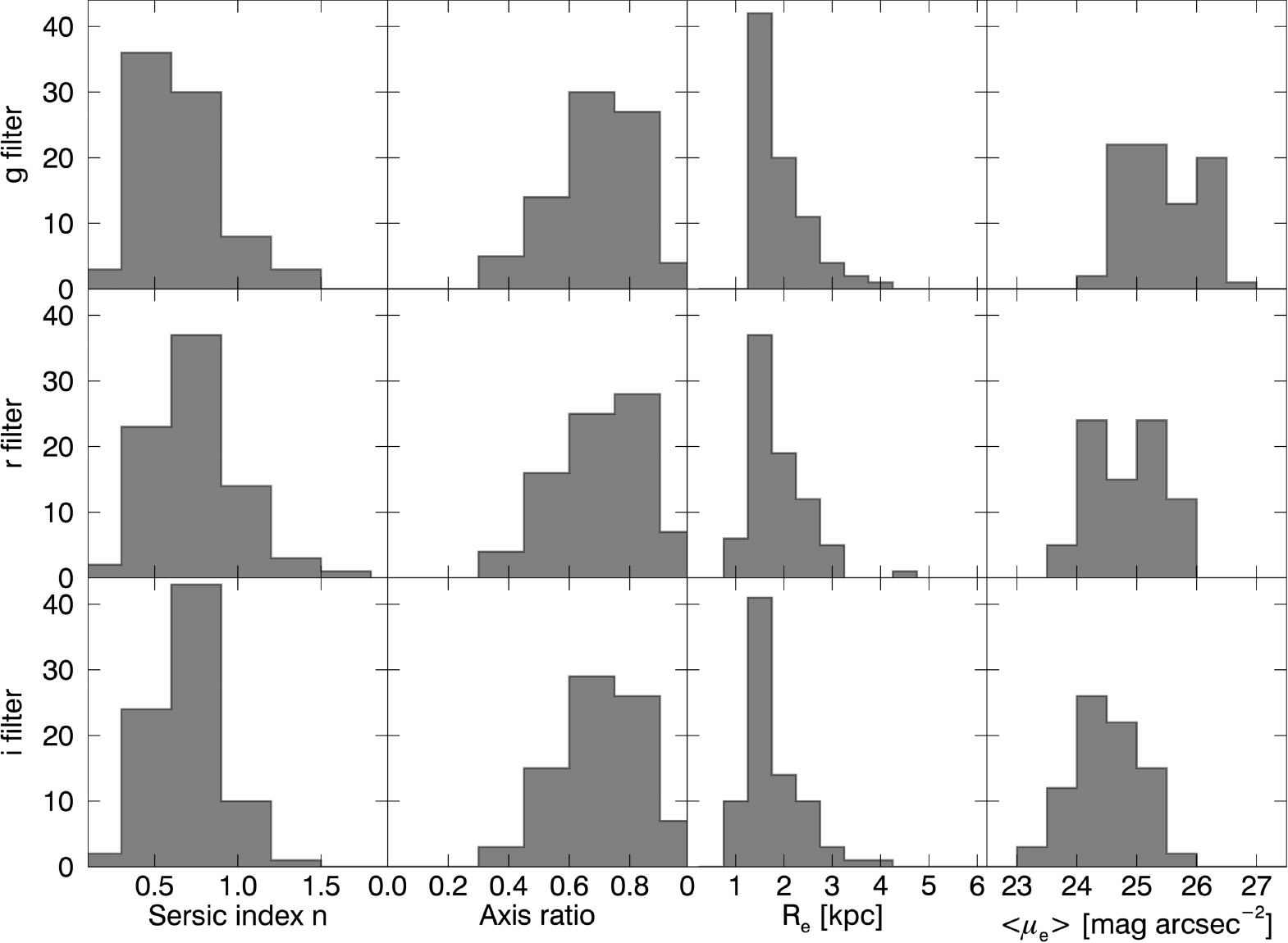}
   \caption{Distribution of structural parameters obtained using a S\'ersic fit to the whole sample of UDGs in the $g$, $r$ and $i$ filters.}
   \label{fig:stat}
\end{figure*}

The final step is to separate galaxies compatible with having the same colour distribution as the spectroscopic sample. For this reason, our final UDG candidates must to share the same colour distribution as the spectroscopic sample in the colour-colour map shown in Fig. \ref{fig:colores}. To do this, we estimate the distance of each of our UDG candidates in the two-dimensional colour space (\textit{g-r}, \textit{r-i}) to the third nearest galaxy with spectroscopy. We select the third nearest object to have a more robust estimation than just using the first. We require that the third object is no further away than 0.2 mag in this colour-colour plane. Additionally we restrict the colour of our UDGs to \textit{g-r} $<$ 0.75 and \textit{r-i} $<$ 0.45 trying to avoid background contaminants from distant red galaxies. Those UDGs satisfying this criterion are shown with red symbols, and those with colours not compatible with the spectroscopic sample are shown with green dots. Our final sample of UDGs compatible with being at the redshift of Abell 168 consists of 80 objects. A representative sample of these UDGs is shown in Fig. \ref{fig:mosaico}. There are 33 galaxies incompatible with being at the cluster distance. This implies a typical contamination level for surveys such as ours that do not have this extra colour of $\sim$30\%. The average colours of the contaminant galaxies are: $<g-r>$ = 0.75 (rms= 0.29) and $<r-i>$ = 0.40 (rms = 0.21).

Despite the fact that we use a third filter to clean our sample of contaminants,  it is clear  that we still should have a level of contamination as a result of foreground and background objects. Not having spectroscopic redshifts it is difficult to estimate accurately the level of this remaining contamination. Our colour-colour, size and surface brightness selection criteria eliminate most of the background interlopers (at least those with z $> $0.1). We have explored in Section \ref{sec:confidence} how much of this contamination could still be in place using the vicinity of the UDG candidates to spectroscopic galaxies in the large-scale structure of Abell 168 and the proximity to spectroscopic galaxies as a proxy.

\section{Properties of the ultra-diffuse galaxy sample}

Once we have selected the sample of UDGs compatible with being at the redshift of cluster Abell 168, we explore the distribution of their structural properties (see Table \ref{tab:parameters}). This is shown in Fig. \ref{fig:stat}.  We show in the figure, for each filter, the distribution of the S\'ersic index $n$, the axial ratio, the circularized effective radius R$_e$ and the average surface brightness {within the effective radius} $<$$\mu_e$$>$. The distribution of all the structural properties is similar in all the bands (see Fig. \ref{fig:appendixtwo}), except $<$$\mu_e$$>$ as expected,  which changes because of the colours of the UDGs. As has been found in previous works \citep[e.g.][]{2015ApJ...807L...2K,2016A&A...590A..20V} the S\'ersic indices $n$ of the UDGs are slightly below 1, with a peak around 0.7 ($<n_{g}>$=0.65, $<n_{r}>$=0.74 and $<n_{i}>$=0.69). Interestingly, almost none of the UDGs has $n>$1.5. Also, the axial ratio of these objects is around 0.7, suggesting a spheroidal shape. Moreover, the number of objects with circularized effective radius larger than 1.5 kpc declines very fast.

\begin{figure}
  \centering
   \includegraphics[width=0.47\textwidth]{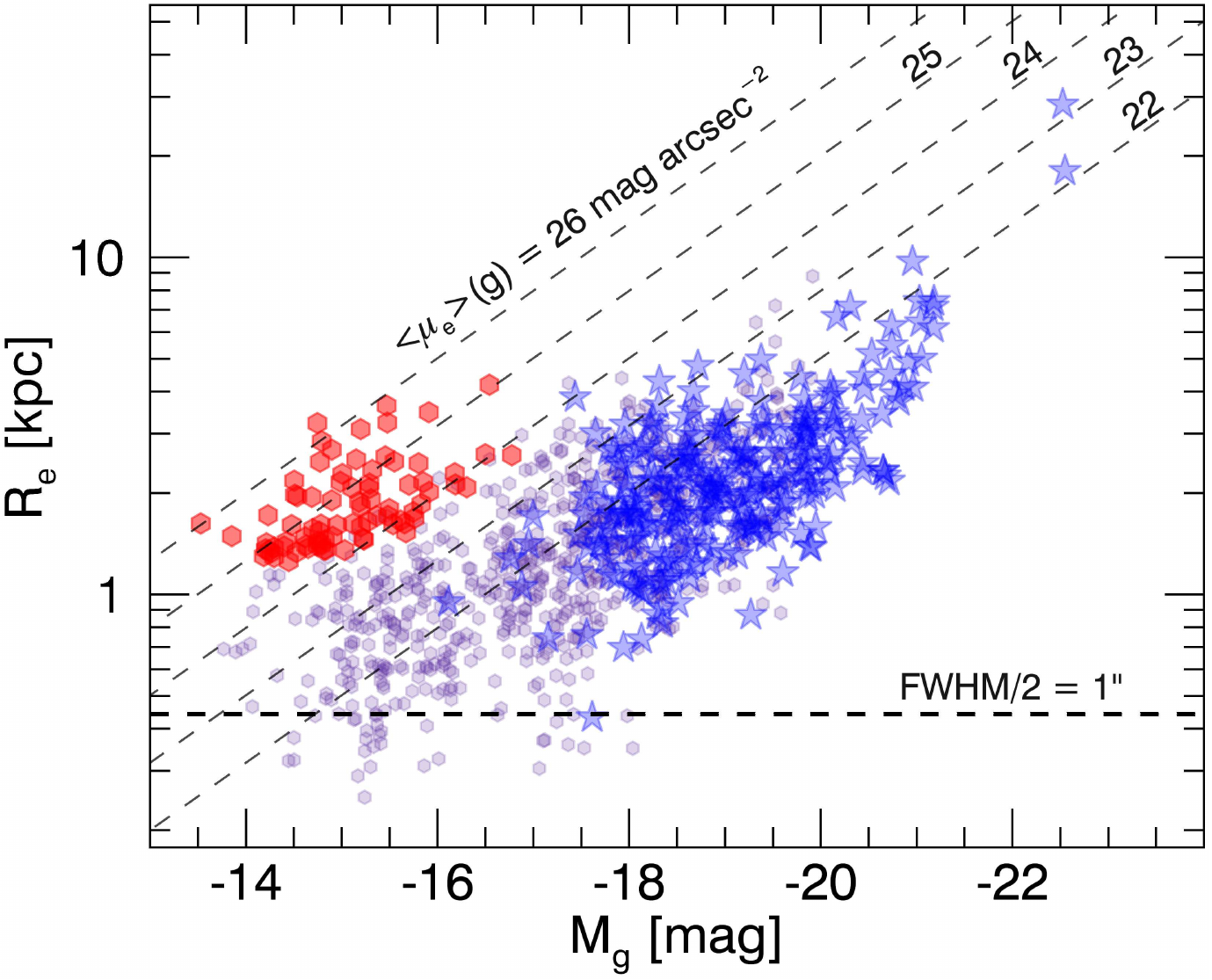}
   
   \caption{Circularized effective radius versus absolute \textit{g}-band magnitude for all the galaxies in the Abell 168 area compatible with being at the cluster redshift. The structural parameters were derived using IMFIT \citep[][]{2015ApJ...799..226E}. Galaxies with spectroscopic redshifts are shown with blue stars, whereas galaxies with photometric redshifts are shown with purple dots. The UDGs are plotted with red hexagons. The approximate value for the {seeing} FHWM of the images is 1 arcsec. In the figure we plot with an horizontal dashed line the value of FWHM/2 which is equivalent to 0.45 kpc at the redshift of the cluster. The inclined dashed lines indicate equal average surface brightness.}

   \label{fig:rmag}
\end{figure}

The apparent \textit{g}-band magnitude of the UDGs within the explored area is 20.5 $<$ $g$ $<$ 22.5 mag. This implies absolute magnitudes of {-16 $\lesssim$ M $_g$ $\lesssim$ -14 mag}. To put our UDGs in context with the rest of the galaxies in the Abell 168 area, we show in Fig. \ref{fig:rmag} the size versus absolute \textit{g}-band magnitude plot for all the galaxies in the field of view at the redshift of the cluster. Galaxies with spectroscopic redshifts are shown with blue stars. We also include galaxies with photometric redshifts (purple dots) using the catalogue from \citet[][]{2012ApJ...747...59R}.  A galaxy with photometric redshift is plotted if its redshift is compatible with being at the redshift of the cluster. We select those that are less than 3$\sigma$ (where $\sigma$ is the photometric redshift error) away from $z$ $=$ 0.045 and with a photometric redshift value of $z$ $<$ 0.15. For the UDGs in our sample, it is not reasonable to use the photometric redshift from \citet[][]{2012ApJ...747...59R}. The reason for this is that the UDGs are significantly fainter (in terms of surface brightness) than the more compact galaxies at the same absolute magnitude. This produces an unacceptable uncertainty of their photometric redshifts.

As has been found previously \citep[see e.g.][]{2015ApJ...807L...2K,2015ApJ...813L..15M}, Fig. \ref{fig:rmag} shows that there is a region where galaxies considered as normal dwarf galaxies and UDGs overlap. This is related to the criterion used for selecting UDGs which uses the central surface brightness (a quantity that it is not directly measured but extrapolated from the S\'ersic fit to the surface brightness distribution of the galaxies). The apparent relation between the size of the UDGs and their absolute magnitude (found here and in previous works) is an artifact produced by the limiting surface brightness of the surveys.

In relation to the colour characterization of our sample, we have a considerable advantage, having three filters compared to the two filters in previous work. \citet[][]{2015ApJ...798L..45V} find an average colour of $<$\textit{g-i}$>$ = 0.8 $\pm$ 0.1, while here we find 0.66 $\pm$ 0.02 (rms = 0.20). According to \citet[][]{2015ApJ...798L..45V}, this colour can be either reproduced by a stellar population with 7 Gyr and [Fe/H] = -1.4 or a population with 4 Gyr and [Fe/H] = -0.8. In both cases, this corresponds to galaxies significantly younger than the most massive galaxies of the cluster. Similarly, \citet[][]{2016A&A...590A..20V} find a typical colour of \textit{g-r} = 0.6 (we find $<$\textit{g-r}$>$ = 0.47 $\pm$ 0.02 with rms = 0.15) which implies an age of 2 Gyr assuming solar metallicity or 6 Gyr with  [Fe/H] = -0.7. Our average colour values are compatible also with relatively modest ages: 3 Gyr ([Fe/H] = -1.3) or 2 Gyr ([Fe/H]= -1.7) \citep[][]{2015MNRAS.449.1177V}. It is worth noting that if we had not corrected the contaminants by their position in the colour-colour map, the average UDG colours would be slightly redder: $<$\textit{g-r}$>$ = 0.55 $\pm$ 0.02 and $<$\textit{r-i} $>=$ 0.25 $\pm$ 0.02.

\begin{figure}
  \centering
   \includegraphics[width=0.47\textwidth]{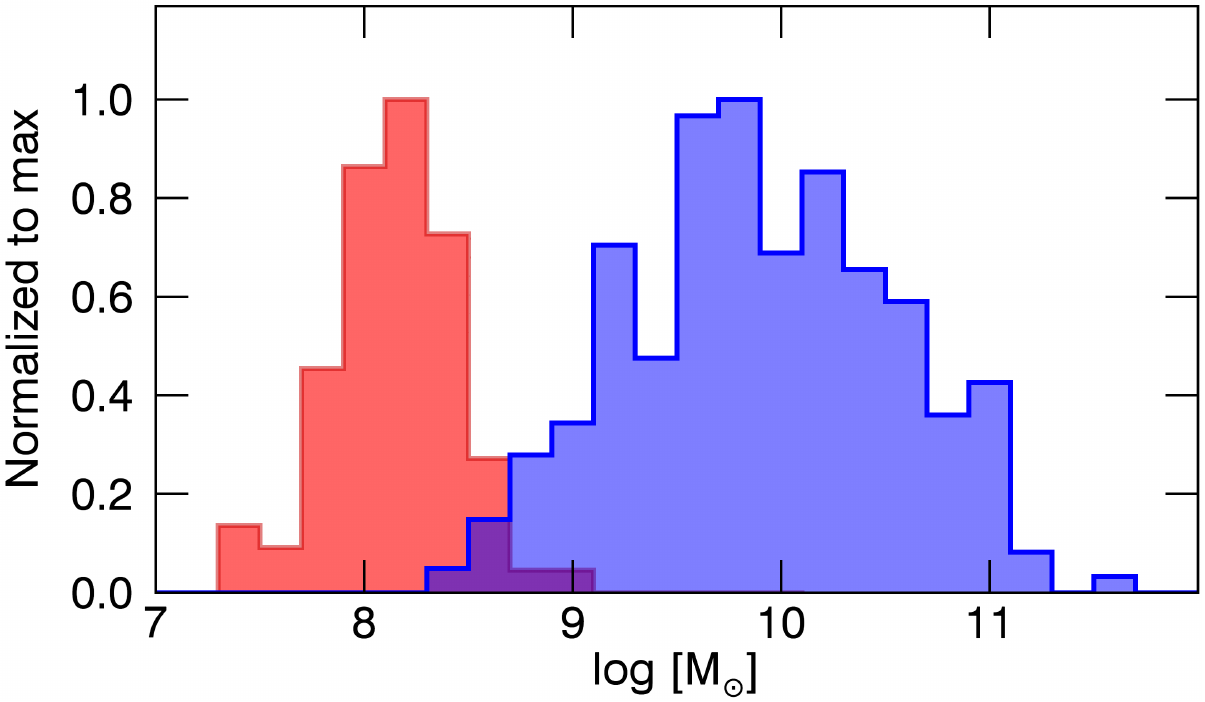}
   
   \caption{Stellar mass distribution of the galaxies in our sample. The red histogram shows the mass distribution of the UDGs whereas the blue histogram shows the distribution of galaxies with spectroscopic redshift {0.037 $< $ $z$ $<$ 0.052}. The histograms have been normalized to peak at 1.}

 \label{fig:histo_masas}
\end{figure}

Finally, we quantify the stellar mass distribution of our UDG galaxies. To do this we take advantage of our colour measurements. In particular, we use \textit{g-r} to determine the mass-to-light ratio in the \textit{r}-band (M/L)$_r$. We follow the method by \citet{2003ApJS..149..289B}. We have used a \citet[][]{2001MNRAS.322..231K} initial mass function (IMF). The stellar mass distribution of our galaxies is shown in Fig. \ref{fig:histo_masas}, with a peak around 10$^8$ M$_\odot$, similar to previous works using surveys with similar depths \citep[see e.g.][]{2016A&A...590A..20V}. As a comparison we show the stellar mass distribution of the galaxies with spectroscopic redshifts 0.037 $<$ $z$ $<$ 0.052.

\section{Spatial distribution of ultra-diffuse galaxies in the large-scale structures}

\begin{figure*}
  \centering
   \includegraphics[width=1.0\textwidth]{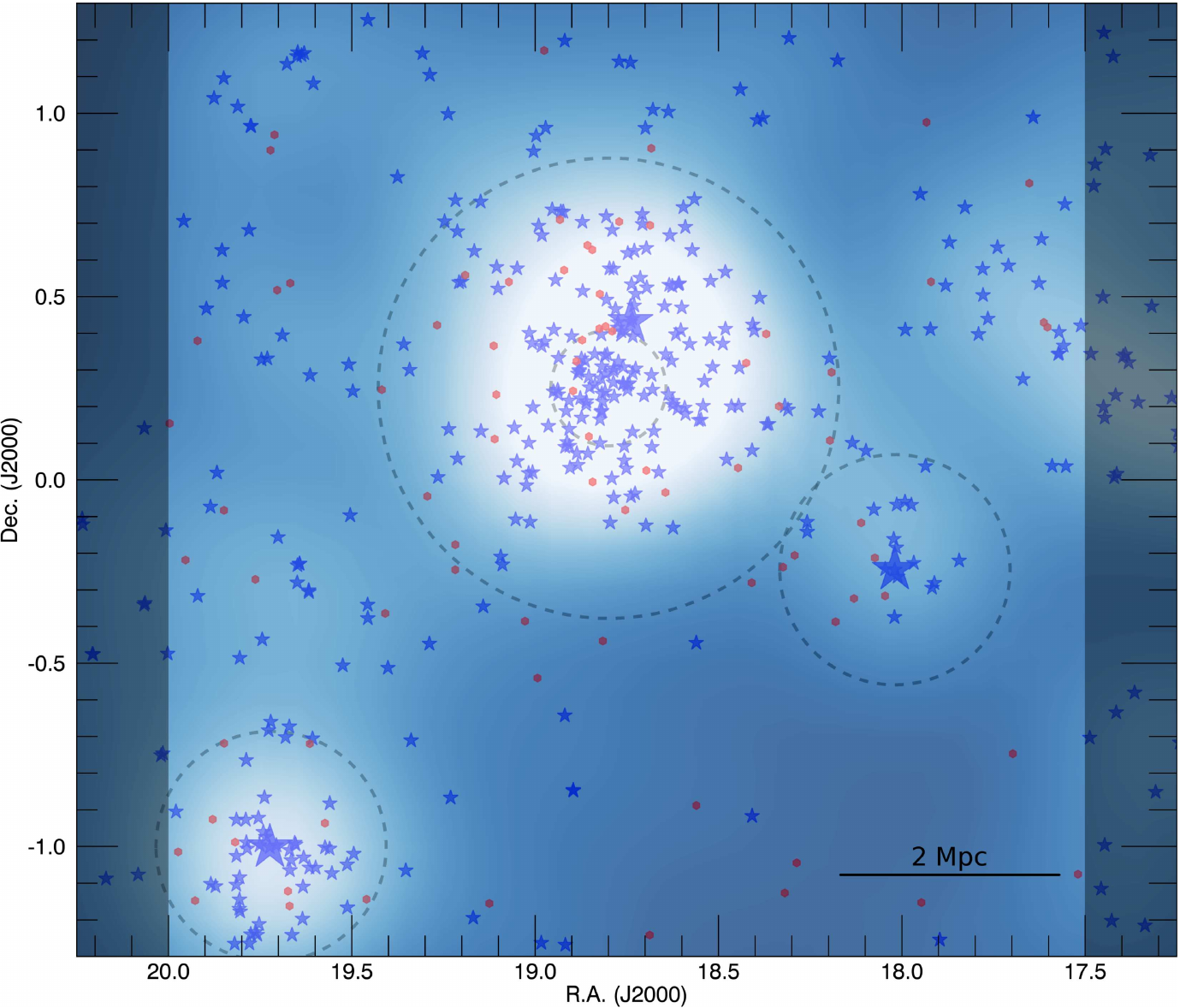}
   
   \caption{Spatial distribution of UDGs in the Abell cluster 168 and its surrounding large-scale structure. The blue stars are the galaxies with spectroscopic redshifts at 0.037 $<$ z $<$ 0.052. Large blue stars correspond to the dominant galaxies in each substructure: UGC 00797 (RA = 18.73996, Dec. = 0.43081, z = 0.04482), UGC 00842 (RA = 19.72338 Dec. = -1.00199, z = 0.04526) and UGC 00753 (RA = 18.01924 Dec. = -0.24512, z = 0.04420). Red dots are the UDGs. The white areas correspond to the density distribution of the galaxies with spectroscopic redshifts. The gray dashed circles enclose  the different zones explored in this paper: the cluster core (R$<$0.5 Mpc) and its outskirts (0.5$<$R$<$2 Mpc), the groups (R$<$1 Mpc) around UGC 00753 and UGC 00842, and filaments (remaining area). The horizontal bar indicates the equivalent size of 2 Mpc at the cluster distance. }

   \label{fig:density_map}
\end{figure*}

The goal of this paper is to explore how UDGs are spatially distributed in a large variety of
environments. Once the final sample of UDGs is selected, we can probe whether they preferentially inhabit a particular region of the complex structure surrounding the Abell Cluster 168. In Fig. \ref{fig:density_map} we show the spatial distribution of the UDGs and the galaxies with spectroscopic redshifts compatible with being at the cluster distance. The size of the area shown is limited by the declination width of the Stripe 82 survey (2.5\degree). We have expanded in the RA direction the area shown by 0.25\degree on each side in order to illustrate how the large scale structure in that direction continues. This is indicated with a darker blue colour. The blue stars are the galaxies with spectroscopic redshifts at 0.037 $<$ $z$ $<$ 0.052. Large blue stars correspond to the dominant galaxies in each prominent substructure: UGC 00797 (RA = 18.73996, Dec. = 0.43081, z = 0.04482), UGC 00842 (RA = 19.72338, Dec. = -1.00199, z = 0.04526) and UGC 00753 (RA = 18.01924, Dec. = -0.24512, z = 0.04420). Red dots are the UDGs. To enhance the visibility of the large-scale structure in  Fig. \ref{fig:density_map} we have estimated the density distribution of the galaxies with spectroscopic redshifts at 0.037 $<$ $z$ $<$ 0.052 using a bin size of 0.1\degree$\times$0.1\degree. Then, we have smoothed the histogram for ease of visualization.

The most prominent structure, located a little above the central part of the plot is the Abell Cluster 168. This cluster is not fully relaxed, with an overdensity of galaxies slightly offset from the position of the main galaxy UGC 00797 \citep[][]{2004ApJ...600..141Y}. The whole central structure has a radius of $\sim$1.5 Mpc. The two other  most conspicuous structures are the fossil group centered around the galaxy UGC 00842 \citep[][]{2010AJ....139..216L} (bottom-left corner of the figure) and the group around UGC 00753. There are other filamentary-like structures, also a big empty region in the bottom-right part of the image. The most remarkable result is that UDGs trace the large-scale structure of spectroscopic galaxies.

\subsection{Effect of interlopers in the spatial distribution of ultra-diffuse galaxies}\label{sec:confidence}

Our selection criteria (colours, size and surface brightness) are constructed to attempt to select all the UDGs at the redshift of the Abell 168 cluster. We try to avoid as much as possible contaminating sources projected in the line of sight. For instance, modeling different stellar population tracks \citep[][]{2015MNRAS.449.1177V} in our colour-colour map, we find that at $z$ $>$ 0.2, the vast majority of these models do not follow our colour-colour selection criteria. Additionally, our selection criterion (r$_e>$ 1.5 kpc at $z$ = 0.045) is equivalent to selecting objects with r$_e>$1.7 arcsec. This value corresponds to the following physical sizes at different redshifts: 3.1 kpc ($z$ = 0.1), 5.6 kpc ($z$ = 0.2) and 7.6 kpc ($z$ = 0.3). Considering that our typical UDG has \textit{g} = 21.5 mag, the absolute \textit{g}-band rest frame (K-corrected) at different redshifts will be -16.7 mag (z = 0.1), -18.3 mag (z = 0.2) and -19.2 mag (z = 0.3). Following Fig. \ref{fig:rmag}, it can be easily seen that galaxies with the above absolute magnitudes and sizes are not  expected to exist. In other words, if they exist, they are not common and consequently, their importance as a source of contamination is expected to be very small. These numbers illustrate that the probability of having in our catalogue of UDGs interlopers with $z$ $>$ 0.1 is very low. For this reason, in what follows, we will concentrate on potential interlopers located in our line of sight up to $z$ = 0.1.

To evaluate the number of potential interlopers within our catalogue of UDGs,  we explore the following idea: we quantify whether a given UDG is more likely to belong to our large-scale structure (at $z$ = 0.045), or whether is more likely to be located in another large-scale structure found in the same field up to $z$ = 0.1. To conduct this task, we assume that the likelihood of a given UDG candidate of belonging to a given structure  is proportional to the number of galaxies with a given spectroscopic redshift in its neighborhood compatible with being part of that structure. On doing this, we are making the following assumption: UDGs are more abundant in the
densest environments. This hypothesis is based on the findings by \citet[][]{2016A&A...590A..20V} and \citet[][]{2016arXiv161008980R}. These authors find a tight correlation between the number of UDGs and the mass of the host structure where they are embedded. This correlation has the following form: N$_{UDGs}$ $\propto$ M$_{Host}^\alpha$ with $\alpha\sim$ 1. Based on this, we estimate the probability of a given UDG in our list to pertain to a given structure as follows: 

\begin{equation} 
P(UDG(z))\propto\sum_{i=1}^{N} M_i (z,M_\star>2\times10^{10}M_\odot,R<0.5 Mpc)
\end{equation}

In other words, we sum the stellar mass of all the galaxies at a given (spectroscopic) redshift around our UDG candidate with M$_\star$ $>$ 2 $\times$ 10$^{10}$M$_\odot$  and within a projected radial distance R $< $ 0.5 Mpc. The reason why we select only spectroscopic galaxies with M$_\star$ $>$ 2 $\times$ 10$^{10}$M$_\odot$ is because this is the completeness stellar mass limit for galaxies up to $z$ = 0.1 in our SDSS spectroscopic sample \citep[see fig. 4 in][]{2014MNRAS.444..682C}. We are also assuming that the total stellar mass contained in the most massive galaxies is a proxy of the total mass of the structure where the UDG candidate is located.

\begin{figure}
  \centering
   \includegraphics[width=0.47\textwidth]{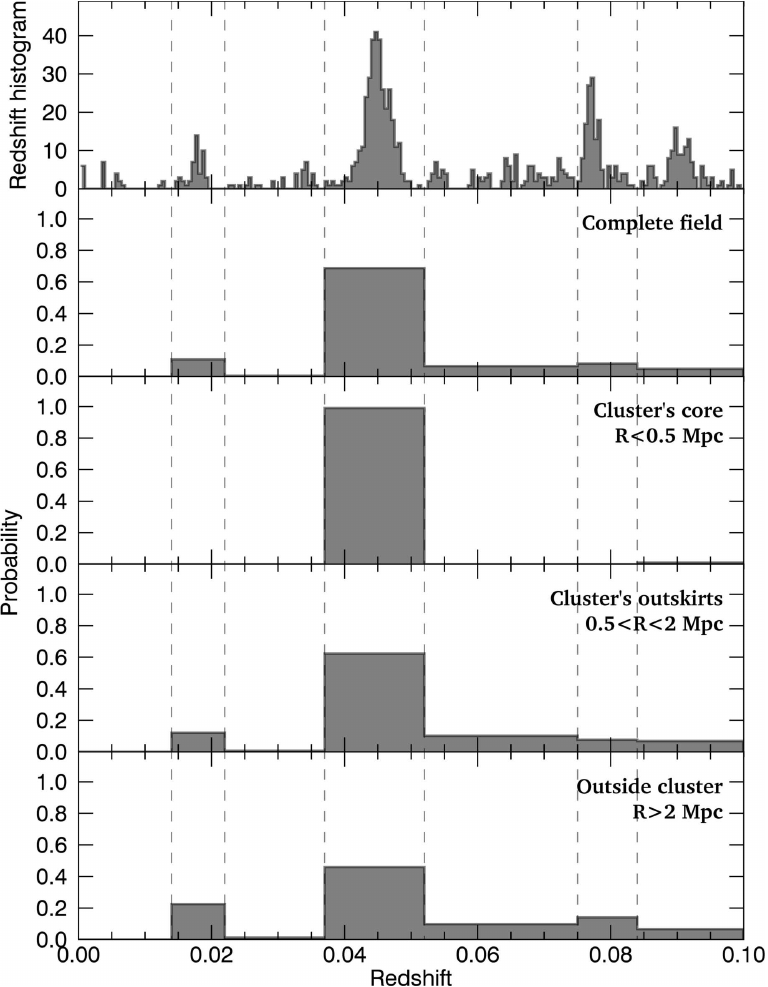}
   
   \caption{Redshift distribution and probability of UDGs to belonging to the different large scale-structures within our field of view up to $z$ = 0.1. The redshift distribution is shown in the upper panel. The other of the panels show the probability of the UDGs being located at a given redshift depending on their position within the field of view. The probabilities of membership of the UDGs in the range of redshift 0.037 $<$ $z$ $<$ 0.052 are: complete field 77 per cent, cluster's core 99 per cent, cluster's outskirts 62 per cent and outside the cluster 46 per cent.}

   \label{fig:contamina}
\end{figure}

To measure the above probability for each UDG candidate, we select different redshift slices following the distribution of the galaxies with spectroscopic redshift in our sample (see Fig.
\ref{fig:contamina}). Once we have estimated the probability for each of our UDG candidates, we sum the probability of the whole sample and we create the probability distribution of the UDGs to belong to different redshift slices within our field of view up to $z$ = 0.1. The result of this analysis is shown in the second row of Fig. \ref{fig:contamina}. This figure shows that 73\% of our UDG candidates are more likely to be located at the large-scale structure at the redshift of the cluster (i.e. $z$ = 0.045). Naturally, the probability is not the same depending on where the UDG candidate is located. For this reason, we have repeated this exercise for different subsamples of our UDG galaxies accounting for their position within our field of view. For the UDGs located in the cluster's core (i.e. in the inner R $<$ 0.5 Mpc), the probability of pertaining to the cluster itself is 99 per cent (in other words, the contamination in this region is negligible). For the region located in the outskirts of the cluster (0.5 $<$ R $<$ 2 Mpc) the probability of UDGs being placed there is 62 per cent. Finally, for the galaxies well outside the cluster the probability of being located there is 46 per cent. It is worth noting that our method is biased against potential UDGs located in the lowest density environments. So, particularly in the regions beyond the cluster's core, our results are likely to be lower limits of the correct value.

Based on the above analysis we can estimate what would be the final number of UDGs in the different structures at $z$ = 0.045. For the cluster's core we will have (after removing the contamination): 5 UDGs, for the cluster's outskirts 17 UDGs and for the large scale structure surrounding the cluster 22 UDGs. This translates into the following percentage of UDG galaxies depending on the structure where they are located: 11 $\pm$ 5 per cent (cluster's core), 39 $\pm$ 9 per cent (cluster's outskirts) and 50 $\pm$ 11 per cent (outside the cluster).

\subsection{Environment of ultra-diffuse galaxies}

\begin{figure}
  \centering
   \includegraphics[width=0.47\textwidth]{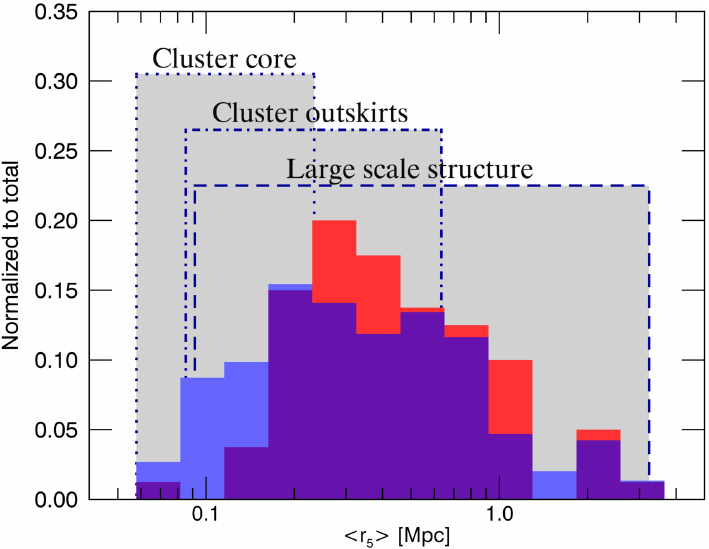}
   
   \caption{Average distance of the galaxies to their first five neighbors $<r_5>$, a proxy for  local
   density (see text for details). The histogram showing the average local density of UDGs is plotted in
   red. The blue histogram shows the same information for galaxies with spectroscopic redshifts
   0.037$<$z$<$0.052 {in the cluster field.}}

   \label{fig:density}
\end{figure}

We characterize the environment around each UDG in our sample by determining the average distance to its first five neighbours with stellar masses $>$ 10$^{9.5}$ M$_\odot$: $<$r$_5>$.  The choice of mass limit is motivated by the stellar mass distribution shape of the galaxies with spectroscopy in Fig. \ref{fig:histo_masas}, which suggests that the spectroscopic sample is complete at $z$ = 0.045 for stellar masses above this value.  In addition, we take the average distance to the closest five neighbors to have a robust estimation (i.e. not strongly affected by shot noise) of the typical distance to the surrounding galaxies. We assume that both UDGs and galaxies with spectroscopic redshifts are all at the same comoving radial distance (i.e. $z$ = 0.045). The result of our analysis is shown in Fig. \ref{fig:density}. For comparison, we also show the local density distribution of all the galaxies with spectroscopic redshifts in the cluster (i.e. also including those with $<$ 10$^{9.5}$ M$_\odot$) and its surrounding large-scale structure. To contextualize the meaning of our local densities, we also show with vertical lines the range in density found for spectroscopic galaxies placed in the core of the Abell Cluster 168, in the cluster outskirts and in the large-scale structure that surrounds it.

The result of this analysis is shown in Fig. \ref{fig:density}. We caution the reader to avoid overinterpreting this figure as the effect of the interlopers has not been corrected. Nevertheless, there is a lack of UDGs in the densest regions of our field as compared with the spectroscopic sample, which cannot be explained as an effect of the false projections. This absence of UDGs in the innermost part of the cluster is in agreement with theoretical and observational work by \citet[][]{2015MNRAS.452..937Y,2016A&A...590A..20V} respectively. Additionally, we have checked the variation of the properties of the UDGs as a function of the density ($<$r$_5>$) in Appendix \ref{sec:dens}. That analysis, which should be considered as a tentative because of the presence of interlopers, can give us some clues about the relationship of UDGs with their environment and it can provide a basis for future work.

The number densities of UDGs in each of the structures are (after removing contamination): 6 $\pm$ 3 Mpc$^{-2}$ in the cluster core, 1.4 $\pm $ 0.4 Mpc$^{-2}$ in the cluster outskirts and 0.4 $\pm$ 0.1 Mpc$^{-2}$ in the large-scale structure. The contributions in stellar mass of the UDGs to the total stellar mass above 3 $\times$ 10$^{9}$ M$_\odot$ in each substructure are: $\sim$0.06 per cent in the cluster core, $\sim$0.08 per cent in the outskirt of the cluster and $\sim$0.10 per cent in the external large scale structure. The contribution of UDGs to the total stellar mass in the cluster is $\sim$0.10 per cent, a number similar (within a factor of 2) to the one (i.e. 0.2 per cent) reported by \citet[][]{2016A&A...590A..20V}.

\section{Ultra-diffuse and dwarf galaxies}

The global distribution of the UDGs in the large-scale structure, shown in Fig. \ref{fig:density_map}, as well as the relationship between their colour and local density suggest a scenario where UDGs are not significantly different, as a population, from dwarf galaxies with stellar masses around 10$^9$ M$_{\sun}$ (i.e. similar to the LMC). In fact, we have explored whether the colours and spatial distribution of UDGs are similar to dwarf galaxies (1 $<$ M$_{\star}<$ 3 $\times$ 10$^{9}$M$_{\sun}$) with spectroscopic redshift in our area\footnote{Ideally, we would like to perform this analysis with dwarf galaxies with lower masses, i.e. closer to the UDGs masses, however, the very small number of spectroscopic objects with such low mass prevents us from doing this. For this reason, we have compromised, and we explore the spatial distribution of 10$^9$ M$_{\sun}$ dwarf galaxies.}. For these galaxies we find the following colours: $<$g-r$>$ = 0.48 $\pm$ 0.02 and $<$r-i$>$ = 0.25 $\pm$ 0.01, very similar to the colours we obtain for the UDGs. Their spatial distribution is also compatible to the UDGs. We find 9 $\pm$ 5 per cent in the cluster core region, 34 $\pm$ 9 per cent in the outskirts of the cluster and 57 $\pm$ 8 per cent in the large-scale structure. carrying out exactly the same analysis but for galaxies with $\sim$5 $\times$ 10$^{10}$M$_{\sun}$ (i.e. Milky Way-like objects), we find:  $<$g-r$>$ = 0.79 $\pm$ 0.02 and $<$r-i$>$ = 0.38 $\pm$ 0.01 and the following number densities: 25 $\pm$ 6\ per cent in the cluster core region, 37 $\pm$ 7 per cent in the outskirt of the cluster and 38 $\pm$ 6 per cent in the large-scale structure. In other words, L$_\star$ galaxies are much redder than UDGs and are also spatially distributed differently than dwarf galaxies and UDGs. It is especially noteworthy that in the inner part of the cluster the fraction of L$_\star$ galaxies is much larger compared to UDGs or dwarf galaxies, which seems to favour a scenario where UDGs are more likely to be bona fide dwarf galaxies than failed L$_\star$ galaxies. We expand on this idea in the following section.

\section{Discussion}

Within the standard galaxy formation scheme, very diffuse galaxies are expected to form in initial fluctuations with low density \citep[][]{1980MNRAS.193..189F,1998MNRAS.295..319M} with blue colours, a large amount of gas and disc-like shapes \citep[][]{1994AJ....107..530M}. Moreover, these objects are not supposed to exist in high-density environments as the effect of tides will easily disrupt them \citep[][]{1986ApJ...303...39D,2009A&A...504..807R,2011ApJ...728...74G}. However, UDGs are numerous in clusters, and the more massive the cluster, the more abundantly they can be found
\citep[e.g.][]{2016A&A...590A..20V,2016arXiv161008980R}. There are two possible scenarios to explain these findings, as follows:

\begin{enumerate}

\item In the first scenario, UDGs are a population connected to the cluster environment and they are born and survive in these dense regions because they are embedded in very massive (Milky Way-like) dark matter halos. They will be disrupted only in the very inner regions of the clusters, because of tidal effects. However, they should be common in the outer regions of clusters.

\item In the second scenario, UDGs are regular dwarf galaxies that have been formed in the lowest-density regions of the large-scale structure. They have dark matter halos typical of dwarf galaxies. This means that, if they are accreted to cluster and/or group central regions, they can survive within them, but farther away from the densest regions compared to the first scenario. Once they reach the peripheral regions of the densest structures, unless they have a radial orbit towards the central regions, they can survive for a long time in the outer parts as they are weakly influenced by dynamical friction \citep[][]{2010MNRAS.405.1723S,2015MNRAS.454.2502S}.

\end{enumerate}

Let us explore whether the current observational evidence supports either of the above scenarios. The strongest observational evidence disfavoring that UDGs are failed L$_\star$ galaxies is the evidence by \citet[][]{2016ApJ...819L..20B}, who find that VCC 1287, an UDG in the Virgo Cluster, has a dark matter halo of $\sim$8 $\times$ 10$^{10}$ M$_{\sun}$. This is more than a factor of 10 smaller than the ones expected for L$_\star$ galaxies ($\gtrsim$ 10$^{12}$ M$_{\sun}$). However, this result is just based on one single galaxy and VCC 1287 is a particularly low mass ($\sim$3 $\times$ 10$^{7}$ M$_{\sun}$) UDG (i.e. a factor of $\sim$3 less massive than most of the UDG population explored so far $\sim$10$^{8}$ M$_{\sun}$). Assuming that in this range of masses, M$_{halo}$ will scale proportionality to  M$_\star$, the result of \citet[][]{2016ApJ...819L..20B} will still disfavor the idea that UDGs with M$_\star\sim$ 10$^{8}$ M$_{\sun}$ are failed L$_\star$ galaxies. This is because the mass of its dark matter halo would still be too low by a factor of 3. It is clear then that we urgently need to estimate the dark matter masses of more UDGs\footnote{Recently, \citet[][]{2016ApJ...822L..31P} and \citet[][]{2016arXiv160408024B} using the number of globular clusters as a proxy for the dark matter halo mass of the galaxy have also found that the UDG DF17 inhabits a dark matter halo with $\sim$10$^{11}$ M$_{\sun}$. Also \citet[][]{2016ApJ...828L...6V} have found $\sim$10$^{12}$ M$_{\sun}$ for the DF44 UDG and \citet[][]{2016arXiv161001595A}, analyzing \textit{Huble Space Telescope} imaging of 54 UDG in the Coma Cluster, have found low-mass haloes for this set of UDGs.}. In the meantime, we can evaluate the outcomes of the other observational results.

A good test to support or disfavor either of the above scenarios is to explore the spatial distribution of UDGs. This test is based on the following two assumptions: (i) the sizes of  galaxies are a direct manifestation of the spin parameter of its halo \citep[][]{1998MNRAS.295..319M}; (ii) the spin distribution is not strongly dependent on environment \citep[][]{2016MNRAS.459L..51A}. Based on the first assumption, it follows that UDGs can be either regular dwarf galaxies with high-spin halos or L$_\star$ galaxies with average spin halos. According to the second assumption, the spatial distribution of UDGs will resemble those of the dark matter haloes where they are embedded. Consequently, we have explored in this paper which spatial distribution UDGs resemble most: L$_\star$ galaxies or those of regular dwarf galaxies. In previous sections we have shown that UDGs share the same spatial distributions and colours as regular dwarf galaxies with stellar mass 10$^9$ M$_{\sun}$, while they do not have the same distribution as more massive (L$_\star$-like) objects. Consequently, this could suggest that we can make the observation that UDGs are more likely to be dwarf galaxies than failed L$_\star$ objects. 

Following a similar argument, \citet[][]{2016MNRAS.459L..51A} make the following prediction. "Under the assumption that the spin distribution is not strongly dependent on environment and that these extended discs are capable of forming stars in a similar way when in isolation, our model suggests that an abundant tail of extended galaxies should be ubiquitous in both clusters and in the field". This is, in fact, what we see in this paper: UDGs are not a phenomenon exclusively linked to clusters of galaxies. Note that previously reported UDGs \citep[with the exception of DGSAT I by][]{2016AJ....151...96M} have been found in dense (i.e. cluster) environments. However, we find that UDGs are common outside the clusters, being clearly located around groups and with hints of existence in the less dense structure in our field\footnote{Recent works by \citep[e.g.][]{2016arXiv160801327D,2016A&A...596A..23S,2016ApJ...833..168M,2016arXiv161008980R,2017arXiv170103804T,2017arXiv170107632B} have demonstrated the existence of UDGs in groups and in the field.}.

Both UDGs and L$_\star$ galaxies are found in the same proportions ($\sim$38 per cent) in the outskirts of the Abell Cluster 168. However, they strongly differ in the inner core region of the cluster (25 per cent L$_\star$ galaxies vs 11 percent UDGs) as well as outside clusters (38 per cent L$_\star$ galaxies vs 50 per cent UDGs). This supports the idea that UDGs are not failed L$_\star$ galaxies as they are scarce in the regions where the tides are strong on these objects. If they were sharing the same type of dark matter halos, then they should  better withstand the harsh conditions of the cluster's core\footnote{An important exception is DF44 \citep[][]{2016ApJ...828L...6V}, which appears to have a massive halo when the globular cluster system of the galaxy is analysed. See also \citep[][]{2017MNRAS.464L.110Z}.}.

A final question we want to address is the following: Do we have any evidence favoring UDGs being formed outside clusters and  later being accreted to the cluster periphery? If this were the case, then this would be another hint that the second proposed scenario is more likely than the first one. Clusters of galaxies are very active locations within the large-scale structure of the Universe, with significant continuous accretion of other galaxies by infall of minor substructures occurring with high frequency \citep[e.g.][]{2010MNRAS.406.2267F}. For this reason, UDGs formed outside Abell 168 have to infall to the cluster at some point. This being the case, both the structural and colour properties of UDGs should reflect a gradual change in their values as we approach to the densest regions. \citet[][]{2015MNRAS.452..937Y} suggest that UDGs undergoing tidal fields should show a decrease on their stellar mass, a flattening of their structure (i.e. a decrease of their axial ratio) and a decline in their surface brightness. In this work, we have looked for any hint of such structural transformation of UDGs (see Appendix \ref{sec:dens}). Given the possible presence of interlopers, especially in the lower-density areas, our work does not allow us to obtain a robust analysis. However, assuming these limitations, we have found a decrease in radius, surface brightness and stellar mass with increasing density. Very interestingly, in their recent work \citet{2016arXiv161008980R} found a similar trend using a different environment, of three isolated compact groups. Although these trends cannot be considered robust at this time, and must be confirmed by later and more extensive works, these would point to the progressive transformation of the UDGs by their infall to the cluster. Similarly, \citet[][]{2016MNRAS.459L..51A} predicts: "it can be expected that the isolated counterparts of cluster UDGs should have more clearly discy morphologies, and not appear as red and quenched". In this work, we do not find any clear hint, either in colour or in axial ratio, to support this conclusion\footnote{It is worth noting that our selection criterion for selecting UDGs could be biased against disc-like galaxies for two reasons: it is based on circular R$_e$ (i.e. it avoids selecting the most inclined galaxies) and it is based on central surface brightness (which again means that inclined projections are disfavored).}.

\section{Summary}

In this paper we have explored the properties of UDGs inhabiting the Abell Cluster 168 ($z$ = 0.045) and its rich surrounding large-scale structure. This work represents an important step forward in understanding in which environments UDGs are born and it allows us to address in more detail the ultimate nature of the UDGs. For instance, among others scenarios, we ask whether most of UDGs are regular dwarf galaxies or failed L$_\star$ galaxies. The main conclusions of this observational work can be summarized as follows:

\begin{enumerate}

\item UDGs are found over the whole scale structure defined by the Abell cluster 168 and its surroundings. Our data allow us to confirm the existence of UDGs in the Abell cluster 168 and groups present in our field, showing hints of the presence of UDGs in the filamentary structure, which is the first detection of UDGs outside a cluster of galaxies.

\item  UDGs are distributed (after removal of potential interlopers) as follows:  11 $\pm$ 5 per cent in the cluster's core, 39 $\pm$ 9 per cent in the outskirts of the cluster and 50 $\pm$ 11 per cent in the large-scale structure around the cluster. The number densities of UDGs in each of the substructures are: 6 $\pm$ 3 Mpc$^{-2}$ in the cluster core, 1.4 $\pm$ 0.4 Mpc$^{-2}$ in the cluster outskirts and 0.4 $\pm$ 0.1 Mpc$^{-2}$ in the large-scale structure.

\item The spatial distribution of the UDGs is similar to that found for regular ($\sim$10$^9$ M$_{\sun}$) dwarf  galaxies but significantly different from that of L$_\star$ objects. Under the assumption that the spin distribution is not strongly dependent on the environment this can be understood as favoring the idea that UDGs are dwarf galaxies inhabiting high-spin halos
\citep[][]{2016MNRAS.459L..51A}.

\item The colours of UDGs ($<$g-r$>$ = 0.48 $\pm$ 0.02, $<$r-i$>$ = 0.21 $\pm $0.02) are compatible with the dwarf galaxies ($<$g-r$>$ = 0.48 $\pm$ 0.02, $<$r-i$>$ = 0.25 $\pm$ 0.01) in the analyzed large scale structure. The colours of L$_\star$ galaxies ($<$g-r$>$ = 0.79 $\pm$ 0.02 and $<$r-i$>$ = 0.38 $\pm$ 0.01), much redder, are indicative of the dwarf nature of the UDGs.

\end{enumerate}

\section*{Acknowledgments}

We are grateful to  the referees for their constructive comments. We want to specially thank Mike Beasley for his helpful comments during the development of this work. We thank Juergen Fliri for his careful work on the IAC Stripe 82 Legacy Project. Lee Kelvin is also acknowledged for his help. This research was supported by the Instituto de Astrof\'isica de Canarias. The authors of this paper acknowledge support from grant AYA2013-48226-C3-1-P from the Spanish Ministry of Economy and Competitiveness (MINECO). JR thanks MINECO for financing his PhD through an FPI grant.

\appendix

\section{Properties of UDGs as a function of their environment}\label{sec:dens} 

\begin{figure}
  \centering
   \includegraphics[width=0.47\textwidth]{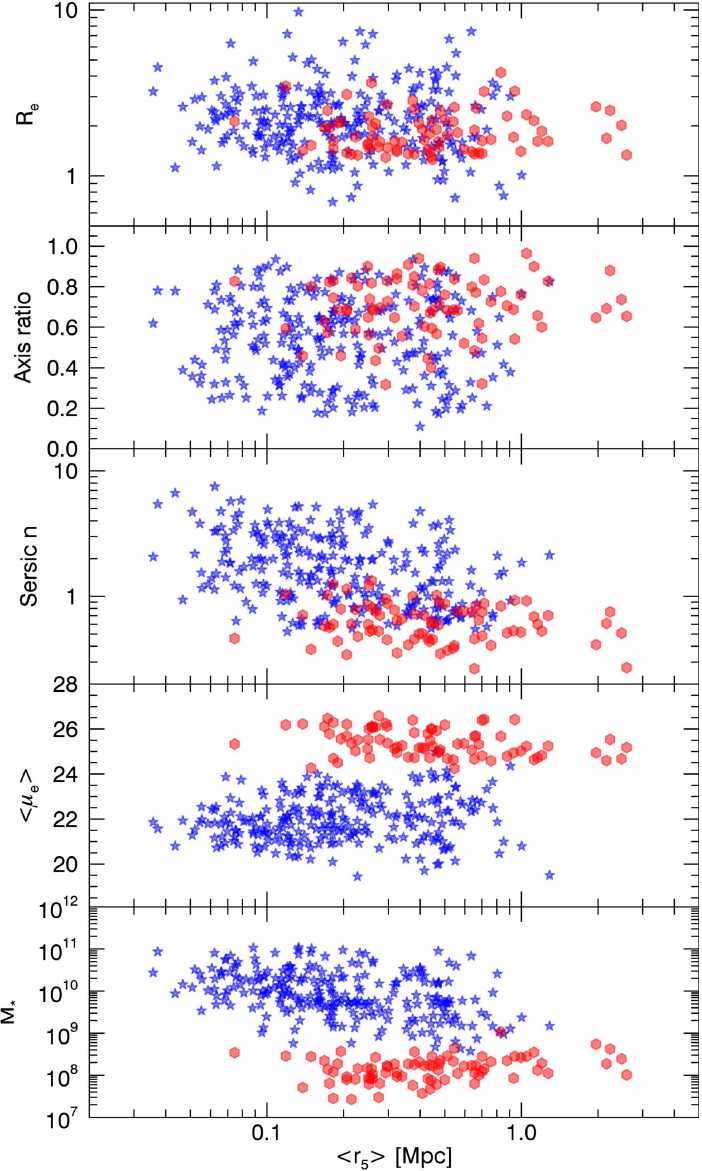}
   
   \caption{Structural properties of the galaxies in our sample (in the \textit{g}-band) versus the average distance of the galaxies to their closest five neighbors with spectroscopic redshift $<r_5>$ (a proxy for the local density): The distribution of the UDGs is in red and the galaxies with spectroscopic redshifts 0.037 $<$ $z$ $<$ 0.052 are in blue.}

   \label{fig:structuraldensity}
\end{figure}

If UDGs were progressively infalling to the cluster center, then it would be expected that during such a process their structural properties would be modified until they become disrupted \citep[][]{2015MNRAS.452..937Y}. To explore any trend between the structural parameters and  local density (see Fig. \ref{fig:structuraldensity}), we separate our sample of UDGs into two groups with similar numbers of galaxies: those with $<r_5>$ $<$ 0.45 Mpc and those with $<r_5>$ $>$ 0.45 Mpc. In Table \ref{tab:density} we show the results of this exercise. We find a dependence on the structural properties of our UDGs sample as a function of the environment. We find a decrease of the stellar mass (by a factor of $\sim$1.5), a decrease of the radius (by a mean of 0.23 kpc), fainter average surface brightness ($\sim$0.4 mag arcsec$^{-2}$), a larger S\'ersic index $n$ (a factor of $\sim$1.1) and a marginal lower axial ratio. These numbers are not corrected by the effect of contaminants as we assume that the interlopers will not modify the average properties of our subsamples.

\begin{figure}
  \centering
   \includegraphics[width=0.47\textwidth]{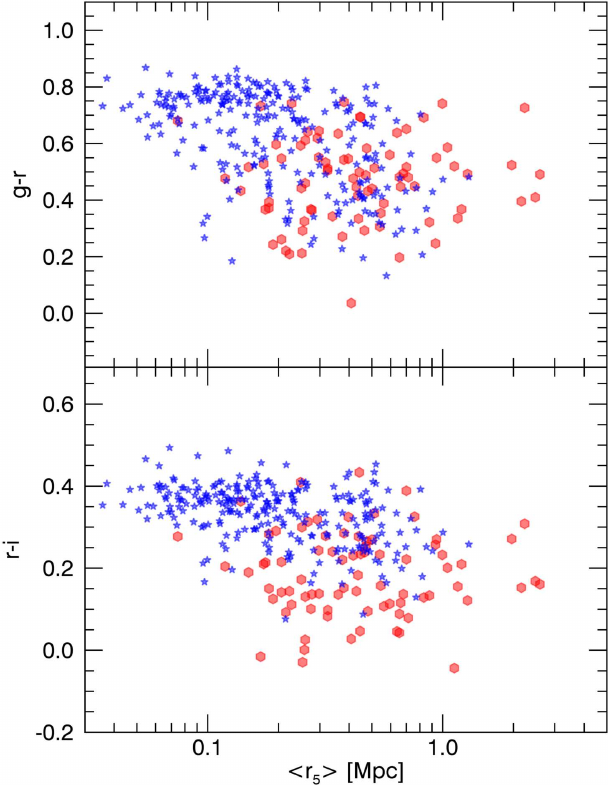}
   
   \caption{Colour versus the average distance of our galaxies to their closest five neighbors with spectroscopic redshift $<r_5>$ (a proxy for the local density). The distribution of the UDGs is shown in red and the galaxies with spectroscopic redshifts 0.037 $<$ z $<$ 0.052 are shown in blue.}

   \label{fig:colordensity}
\end{figure}

Additionally, we explore whether the colour distribution of the UDGs depends on the environment where they reside. In Fig. \ref{fig:colordensity}, we show the colour distribution of the UDGs as a function of their local density as characterized by the average distance to their five closest neighbors $<r_5>$. According to that figure, the colours of UDGs are  independent of their local density.

\begin{table} 
\caption{Average properties of the UDGs as a function of their local density $<r_5>$.}
\begin{tabular}{cccc}
 & & $<r_5>$ $<$ 0.45 Mpc & $<r_5>$ $>$ 0.45 Mpc\\
\hline 
 $<g-r>$     &                            & 0.47 $\pm$ 0.02  & 0.47 $\pm $0.02 \\
 $<r-i>$     &                            & 0.19 $\pm$ 0.02  & 0.18 $\pm $0.02 \\
 $<n>$       &                            & 0.67 $\pm $0.04  & 0.62 $\pm$ 0.04 \\
 $<r_e>$     & (kpc)                      & 1.71 $\pm$ 0.06  & 1.94 $\pm $0.09 \\
 $<\mu_e>$   & (mag arcsec$^{-2}$)           & 25.6 $\pm$ 0.1   & 25.2 $\pm $0.1  \\
 $<b/a>$     &                            & 0.69 $\pm$ 0.02  & 0.71 $\pm$ 0.02 \\
 $<M_\star>$ & ($\times$10$^8$ M$_{\sun}$) &  1.3 $\pm $0.1   & 1.9 $\pm$ 0.2 \\
\hline
\end{tabular}
\label{tab:density}
\end{table}

\section{Robustness of the structural parameters}

\begin{figure}
  \begin{centering}
   \includegraphics[width=0.47\textwidth]{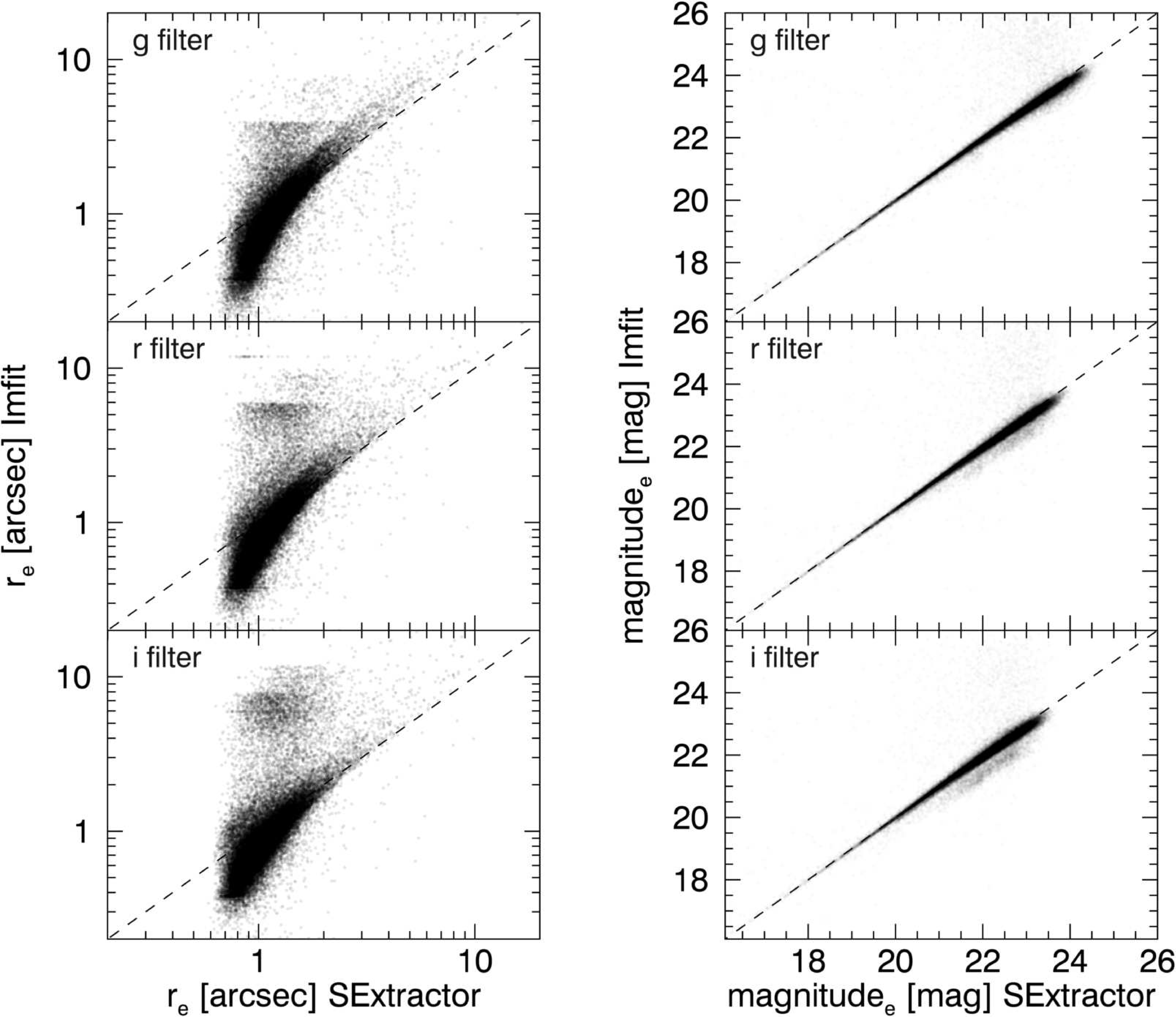}
  
   \caption{Size and apparent magnitude comparison between SExtractor and IMFIT for all the galaxies
   detected in the cluster field.}
  
   \label{fig:appendixone}
\end{centering}
\end{figure}

In this appendix, we compare the structural properties, apparent magnitude and effective radius provided by IMFIT and SExtractor for the whole set of galaxies in our field. This comparison is done for those objects that satisfy the SExtractor detection criteria and the primordial colour cut explained in Section 3. Our results are shown in Fig. \ref{fig:appendixone}. In Table \ref{tab:parameters}, we list the positions and structural properties of the selected UDG candidates in the area surrounding Abell Cluster 168.

There is a very good agreement in all the filters related to the global apparent magnitude of the objects. At the magnitude limit where most of the UDGs are located (20.5 $<$ $g$ $<$ 22.5 mag), the rms between SExtractor and IMFIT is around 0.1 mag. In relation to the effective radius along the semimajor axis, r$_e$, we find that there is a good 1 : 1 relation for those galaxies larger than the typical seeing of the Stripe 82 data ($\sim$1 arcsec). This is as expected, as SExtractor does not take into account the seeing of the data when estimating r$_e$. For this reason, their r$_e$ is overestimated. The typical r$_e$ of the UDGs in our image is $\sim$2 arcsec. At these values, the rms in the size estimation between both codes is 10 per cent.

\begin{figure}
  \centering
   \includegraphics[width=0.47\textwidth]{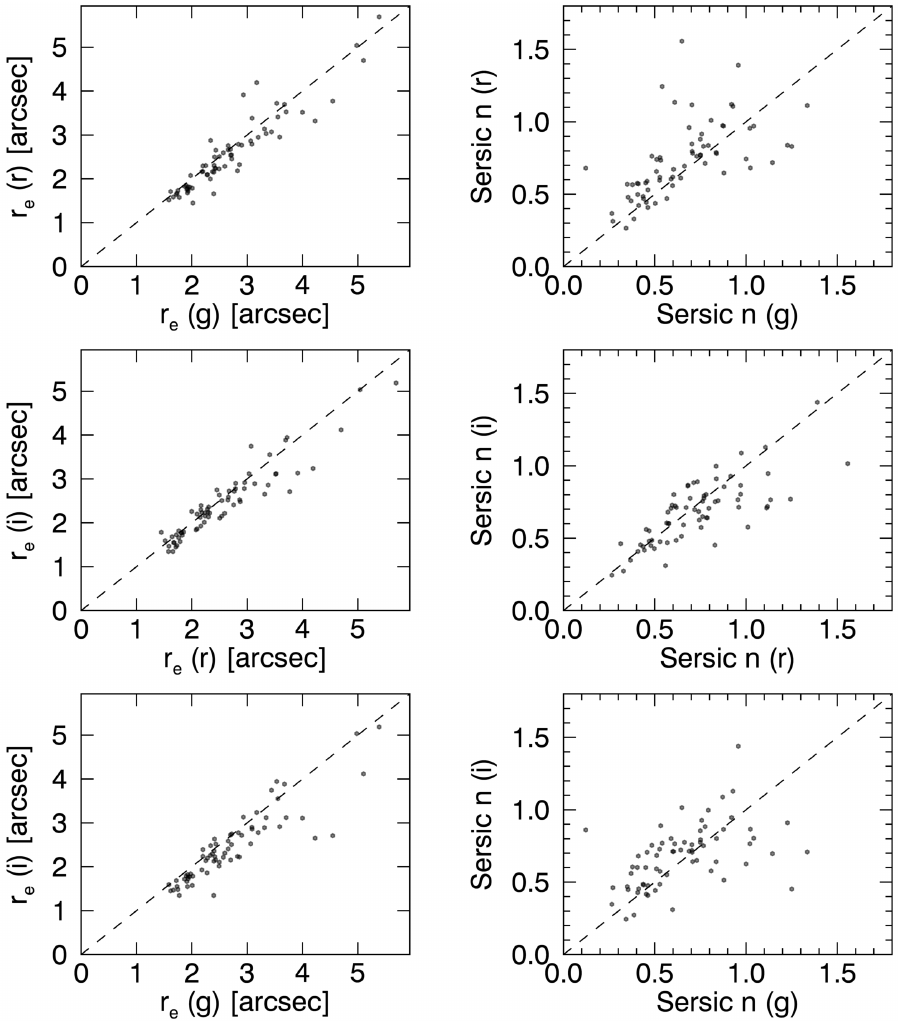}
  
   \caption{Robustness of the structural parameters r$_e$ and $n$ obtained with IMFIT for the set of UDGs explored in this work. The size and
   shape of the UDGs are compared between the different filters: g, r and i.}
  
   \label{fig:appendixtwo}
\end{figure}

In addition, we have checked the robustness of the structural parameters for our UDGs comparing the sizes and S\'ersic indeces among the different filters used. The result of this comparison is shown in Fig. \ref{fig:appendixtwo}. We do find a good agreement between these structural parameters independent of the filter used. As expected, the effective radius along the semimajor axis r$_e$ is estimated more precisely than the S\'ersic index.

\newpage

\onecolumn
\section{Catalogue of UDGs}
\begin{longtable}{c c c c c c c c}
\hline
ID & R.A. & Dec & $\mu_g$(0) & r$_{e}$ & M$_{g}$ &b/a & n\\
  &(J2000)&(J2000)&mag arcsec$^{-2}$& (kpc)&(mag)& &\\
\hline

IAC01&19.9966&0.1540&25.4$^{+0.5}_{-0.5}$&4.4$\pm$0.4&-15.6$\pm$0.1&0.54$\pm$0.08&0.93$\pm$0.23\\
IAC02&19.9733&-1.0147&25.3$^{+0.5}_{-0.4}$&2.5$\pm$0.2&-14.6$\pm$0.1&0.76$\pm$0.11&0.87$\pm$0.22\\
IAC03&19.9533&-0.2186&24.8$^{+0.4}_{-0.3}$&1.7$\pm$0.2&-14.6$\pm$0.1&0.62$\pm$0.09&0.65$\pm$0.16\\
IAC04&19.9272&-1.1476&24.4$^{+0.4}_{-0.4}$&2.0$\pm$0.2&-15.2$\pm$0.1&0.65$\pm$0.10&0.78$\pm$0.19\\
IAC05&19.9205&0.3794&24.4$^{+0.3}_{-0.2}$&1.9$\pm$0.2&-15.6$\pm$0.1&0.73$\pm$0.11&0.41$\pm$0.10\\
IAC06&19.8794&-0.9259&25.0$^{+0.2}_{-0.2}$&1.5$\pm$0.2&-14.5$\pm$0.1&0.86$\pm$0.13&0.35$\pm$0.09\\
IAC07&19.8484&-0.0829&24.6$^{+0.2}_{-0.2}$&1.4$\pm$0.1&-14.8$\pm$0.1&0.94$\pm$0.14&0.27$\pm$0.07\\
IAC08&19.8480&-0.7186&25.5$^{+0.4}_{-0.3}$&2.1$\pm$0.2&-14.3$\pm$0.1&0.40$\pm$0.06&0.63$\pm$0.16\\
IAC09&19.8180&-0.9883&24.1$^{+0.2}_{-0.2}$&1.7$\pm$0.2&-15.7$\pm$0.1&0.80$\pm$0.12&0.38$\pm$0.09\\
IAC10&19.7629&-0.2714&24.8$^{+0.3}_{-0.3}$&2.2$\pm$0.2&-15.3$\pm$0.1&0.78$\pm$0.12&0.51$\pm$0.13\\
IAC11&19.7214&0.8993&24.9$^{+0.2}_{-0.2}$&2.6$\pm$0.3&-15.8$\pm$0.1&0.90$\pm$0.13&0.35$\pm$0.09\\
IAC12&19.7106&0.9414&24.5$^{+0.4}_{-0.4}$&2.1$\pm$0.2&-15.3$\pm$0.1&0.60$\pm$0.09&0.70$\pm$0.18\\
IAC13&19.7034&0.5177&24.4$^{+0.4}_{-0.4}$&1.6$\pm$0.2&-14.7$\pm$0.1&0.80$\pm$0.12&0.75$\pm$0.19\\
IAC14&19.6741&-1.1219&25.2$^{+0.3}_{-0.2}$&1.5$\pm$0.1&-14.1$\pm$0.1&0.80$\pm$0.12&0.48$\pm$0.12\\
IAC15&19.6678&0.5367&24.5$^{+0.3}_{-0.2}$&2.2$\pm$0.2&-15.7$\pm$0.1&0.55$\pm$0.08&0.48$\pm$0.12\\
IAC16&19.6697&-1.1622&25.7$^{+0.3}_{-0.3}$&2.4$\pm$0.2&-14.6$\pm$0.1&0.68$\pm$0.10&0.54$\pm$0.14\\
IAC17&19.6144&-0.7196&24.3$^{+0.3}_{-0.3}$&1.7$\pm$0.2&-15.2$\pm$0.1&0.74$\pm$0.11&0.53$\pm$0.13\\
IAC18&19.5731&-0.9363&25.8$^{+0.3}_{-0.2}$&3.0$\pm$0.3&-15.0$\pm$0.1&0.83$\pm$0.12&0.45$\pm$0.11\\
IAC19&19.4598&-1.1439&25.5$^{+0.4}_{-0.3}$&2.8$\pm$0.3&-15.0$\pm$0.1&0.59$\pm$0.09&0.61$\pm$0.15\\
IAC20&19.4180&0.2460&24.8$^{+0.3}_{-0.3}$&2.1$\pm$0.2&-15.2$\pm$0.1&0.94$\pm$0.14&0.50$\pm$0.13\\
IAC21&19.4088&-0.3643&24.0$^{+0.3}_{-0.2}$&2.8$\pm$0.3&-16.8$\pm$0.1&0.84$\pm$0.13&0.40$\pm$0.10\\
IAC22&19.2940&-0.0448&24.4$^{+0.2}_{-0.2}$&2.6$\pm$0.3&-16.2$\pm$0.1&0.69$\pm$0.10&0.39$\pm$0.10\\
IAC23&19.2672&0.4220&24.5$^{+0.2}_{-0.2}$&2.1$\pm$0.2&-15.7$\pm$0.1&0.68$\pm$0.10&0.37$\pm$0.09\\
IAC24&19.2176&-0.1766&24.6$^{+0.4}_{-0.4}$&2.0$\pm$0.2&-14.9$\pm$0.1&0.52$\pm$0.08&0.77$\pm$0.19\\
IAC25&19.2178&-0.2451&25.7$^{+0.4}_{-0.4}$&2.4$\pm$0.2&-14.3$\pm$0.1&0.32$\pm$0.05&0.75$\pm$0.19\\
IAC26&19.1906&0.5585&24.4$^{+0.3}_{-0.2}$&1.6$\pm$0.2&-15.1$\pm$0.1&0.70$\pm$0.11&0.46$\pm$0.11\\
IAC27&19.1249&-1.1553&24.4$^{+0.3}_{-0.3}$&2.7$\pm$0.3&-16.1$\pm$0.1&0.71$\pm$0.11&0.51$\pm$0.13\\
IAC28&19.1129&0.3658&25.5$^{+0.3}_{-0.2}$&2.2$\pm$0.2&-14.6$\pm$0.1&0.45$\pm$0.07&0.46$\pm$0.12\\
IAC29&19.1103&0.1114&24.4$^{+0.4}_{-0.3}$&1.5$\pm$0.2&-14.7$\pm$0.1&0.84$\pm$0.13&0.64$\pm$0.16\\
IAC30&19.1059&0.2328&24.4$^{+0.3}_{-0.2}$&2.0$\pm$0.2&-15.5$\pm$0.1&0.81$\pm$0.12&0.44$\pm$0.11\\
IAC31&19.0715&0.5402&25.2$^{+0.4}_{-0.4}$&2.7$\pm$0.3&-15.1$\pm$0.1&0.85$\pm$0.13&0.73$\pm$0.18\\
IAC32&19.0276&-0.3854&24.5$^{+0.3}_{-0.3}$&2.1$\pm$0.2&-15.4$\pm$0.1&0.69$\pm$0.10&0.53$\pm$0.13\\
IAC33&18.9934&-0.5405&24.6$^{+0.4}_{-0.4}$&1.8$\pm$0.2&-14.8$\pm$0.1&0.83$\pm$0.12&0.70$\pm$0.18\\
IAC34&18.9753&1.1713&25.4$^{+0.3}_{-0.2}$&2.7$\pm$0.3&-15.2$\pm$0.1&0.48$\pm$0.07&0.46$\pm$0.11\\
IAC35&18.9325&0.7096&24.2$^{+0.3}_{-0.2}$&2.5$\pm$0.3&-16.3$\pm$0.1&0.69$\pm$0.10&0.44$\pm$0.11\\
IAC36&18.9211&0.5722&25.3$^{+0.6}_{-0.5}$&3.2$\pm$0.3&-14.9$\pm$0.1&0.81$\pm$0.12&1.00$\pm$0.25\\
IAC37&18.8968&0.2427&25.0$^{+0.6}_{-0.5}$&4.5$\pm$0.5&-15.8$\pm$0.1&0.59$\pm$0.09&1.04$\pm$0.26\\
IAC38&18.8872&0.3236&25.0$^{+0.3}_{-0.2}$&2.3$\pm$0.2&-15.3$\pm$0.1&0.83$\pm$0.12&0.46$\pm$0.12\\
IAC39&18.8716&0.3812&24.8$^{+0.6}_{-0.6}$&3.7$\pm$0.4&-15.4$\pm$0.1&0.69$\pm$0.10&1.14$\pm$0.29\\
IAC40&18.8571&0.6398&24.5$^{+0.5}_{-0.4}$&1.7$\pm$0.2&-14.6$\pm$0.1&0.73$\pm$0.11&0.88$\pm$0.22\\
IAC41&18.8534&0.1183&25.3$^{+0.6}_{-0.5}$&3.3$\pm$0.3&-14.9$\pm$0.1&0.57$\pm$0.09&1.02$\pm$0.26\\
IAC42&18.8441&0.6275&25.4$^{+0.4}_{-0.4}$&3.1$\pm$0.3&-15.2$\pm$0.1&0.44$\pm$0.07&0.76$\pm$0.19\\
IAC43&18.8432&-0.0056&24.9$^{+0.4}_{-0.3}$&1.5$\pm$0.1&-14.2$\pm$0.1&0.84$\pm$0.13&0.60$\pm$0.15\\
IAC44&18.8248&0.4125&25.5$^{+0.2}_{-0.2}$&1.7$\pm$0.2&-14.2$\pm$0.1&0.68$\pm$0.10&0.34$\pm$0.09\\
IAC45&18.8234&0.5071&24.3$^{+0.4}_{-0.4}$&2.2$\pm$0.2&-15.5$\pm$0.1&0.59$\pm$0.09&0.79$\pm$0.20\\
IAC46&18.8153&-0.4392&24.2$^{+0.3}_{-0.3}$&2.0$\pm$0.2&-15.5$\pm$0.1&0.66$\pm$0.10&0.60$\pm$0.15\\
IAC47&18.8077&0.4186&25.3$^{+0.3}_{-0.3}$&2.1$\pm$0.2&-14.6$\pm$0.1&0.82$\pm$0.12&0.57$\pm$0.14\\
IAC48&18.7897&0.4053&25.6$^{+0.4}_{-0.4}$&2.1$\pm$0.2&-14.1$\pm$0.1&0.46$\pm$0.07&0.71$\pm$0.18\\
IAC49&18.7707&0.7043&24.2$^{+0.3}_{-0.3}$&1.7$\pm$0.2&-15.2$\pm$0.1&0.74$\pm$0.11&0.60$\pm$0.15\\
IAC50&18.7542&-0.0821&24.7$^{+0.7}_{-0.6}$&1.8$\pm$0.2&-13.7$\pm$0.1&0.83$\pm$0.12&1.25$\pm$0.31\\
IAC51&18.6968&0.0255&24.8$^{+0.5}_{-0.4}$&3.1$\pm$0.3&-15.7$\pm$0.1&0.46$\pm$0.07&0.79$\pm$0.20\\
IAC52&18.6864&0.6942&25.6$^{+0.3}_{-0.3}$&2.5$\pm$0.2&-14.6$\pm$0.1&0.63$\pm$0.09&0.58$\pm$0.15\\
IAC53&18.6833&0.9045&24.5$^{+0.3}_{-0.3}$&1.7$\pm$0.2&-15.0$\pm$0.1&0.88$\pm$0.13&0.59$\pm$0.15\\
IAC54&18.6877&-1.2416&24.7$^{+0.3}_{-0.2}$&3.2$\pm$0.3&-16.4$\pm$0.1&0.65$\pm$0.10&0.41$\pm$0.10\\
IAC55&18.6447&-0.0344&25.2$^{+0.5}_{-0.5}$&2.3$\pm$0.2&-14.3$\pm$0.1&0.32$\pm$0.05&0.96$\pm$0.24\\
IAC56&18.5604&-0.8880&24.8$^{+0.4}_{-0.4}$&2.6$\pm$0.3&-15.3$\pm$0.1&0.88$\pm$0.13&0.75$\pm$0.19\\
IAC57&18.4465&0.0327&25.5$^{+0.1}_{-0.1}$&1.7$\pm$0.2&-14.4$\pm$0.1&0.67$\pm$0.10&0.12$\pm$0.03\\
IAC58&18.4244&0.3189&24.5$^{+0.7}_{-0.6}$&3.0$\pm$0.3&-15.2$\pm$0.1&0.56$\pm$0.08&1.23$\pm$0.31\\
IAC59&18.4093&-0.2806&24.6$^{+0.3}_{-0.3}$&1.6$\pm$0.2&-14.8$\pm$0.1&0.76$\pm$0.11&0.52$\pm$0.13\\
IAC60&18.3700&0.3980&25.3$^{+0.5}_{-0.4}$&2.1$\pm$0.2&-14.3$\pm$0.1&0.69$\pm$0.10&0.81$\pm$0.20\\
IAC61&18.3349&0.2009&26.2$^{+0.3}_{-0.3}$&2.1$\pm$0.2&-13.8$\pm$0.1&0.50$\pm$0.07&0.53$\pm$0.13\\
IAC62&18.3193&-1.1266&25.1$^{+0.2}_{-0.2}$&1.7$\pm$0.2&-14.7$\pm$0.1&0.65$\pm$0.10&0.27$\pm$0.07\\
IAC63&18.3238&-0.2380&24.8$^{+0.5}_{-0.4}$&4.8$\pm$0.5&-16.5$\pm$0.1&0.78$\pm$0.12&0.84$\pm$0.21\\
IAC64&18.2922&-0.2059&25.5$^{+0.5}_{-0.4}$&3.5$\pm$0.4&-15.1$\pm$0.1&0.83$\pm$0.13&0.88$\pm$0.22\\
IAC65&18.2867&-1.0447&24.3$^{+0.3}_{-0.3}$&2.3$\pm$0.2&-15.9$\pm$0.1&0.74$\pm$0.11&0.51$\pm$0.13\\
IAC66&18.1960&0.1074&24.5$^{+0.3}_{-0.2}$&1.4$\pm$0.1&-14.8$\pm$0.1&0.91$\pm$0.14&0.41$\pm$0.10\\
IAC67&18.1917&0.2934&24.8$^{+0.6}_{-0.5}$&2.5$\pm$0.3&-14.8$\pm$0.1&0.57$\pm$0.09&1.02$\pm$0.26\\
IAC68&18.1807&-0.3870&24.8$^{+0.5}_{-0.4}$&2.9$\pm$0.3&-15.4$\pm$0.1&0.77$\pm$0.12&0.84$\pm$0.21\\
IAC69&18.1305&-0.3235&24.9$^{+0.3}_{-0.2}$&1.5$\pm$0.2&-14.4$\pm$0.1&0.68$\pm$0.10&0.45$\pm$0.11\\
IAC70&18.1110&-0.1171&24.7$^{+0.4}_{-0.3}$&2.1$\pm$0.2&-15.1$\pm$0.1&0.56$\pm$0.08&0.66$\pm$0.17\\
IAC71&18.0735&-0.2127&24.2$^{+0.5}_{-0.4}$&1.6$\pm$0.2&-14.7$\pm$0.1&0.90$\pm$0.13&0.86$\pm$0.22\\
IAC72&18.0456&-0.3164&24.4$^{+0.7}_{-0.7}$&4.0$\pm$0.4&-15.7$\pm$0.1&0.82$\pm$0.12&1.34$\pm$0.33\\
IAC73&17.9467&-1.1528&24.1$^{+0.4}_{-0.3}$&2.0$\pm$0.2&-15.7$\pm$0.1&0.69$\pm$0.10&0.61$\pm$0.15\\
IAC74&17.9327&0.9754&24.4$^{+0.3}_{-0.3}$&2.4$\pm$0.2&-15.8$\pm$0.1&0.60$\pm$0.09&0.53$\pm$0.13\\
IAC75&17.9204&0.5405&24.1$^{+0.5}_{-0.5}$&1.8$\pm$0.2&-15.0$\pm$0.1&0.71$\pm$0.11&0.89$\pm$0.22\\
IAC76&17.6973&-0.7469&24.2$^{+0.5}_{-0.5}$&2.4$\pm$0.2&-15.4$\pm$0.1&0.96$\pm$0.14&0.92$\pm$0.23\\
IAC77&17.6523&0.8092&24.6$^{+0.4}_{-0.4}$&2.2$\pm$0.2&-15.1$\pm$0.1&0.68$\pm$0.10&0.75$\pm$0.19\\
IAC78&17.6129&0.4294&24.9$^{+0.4}_{-0.4}$&1.9$\pm$0.2&-14.6$\pm$0.1&0.70$\pm$0.10&0.70$\pm$0.18\\
IAC79&17.6030&0.4164&25.0$^{+0.3}_{-0.3}$&1.9$\pm$0.2&-14.7$\pm$0.1&0.65$\pm$0.10&0.60$\pm$0.15\\
IAC80&17.5198&-1.0757&24.0$^{+0.4}_{-0.3}$&2.3$\pm$0.2&-15.9$\pm$0.1&0.90$\pm$0.13&0.69$\pm$0.17\\

\caption{Position and structural properties of the selected UDG candidates in the area surrounding  Abell
Cluster 168. The structural parameters were derived using IMFIT \citep[][]{2015ApJ...799..226E}. In the table, the effective
radius provided r$_e$ is the value along the semimajor axis.}
\label{tab:parameters}
\end{longtable}

\twocolumn

\end{document}